\documentclass{aastex6}

\begin{document}

\newpage

\title{An alternative to the $\Lambda$CDM  model: the case of scale invariance}


\author{Andre Maeder}
\affil{Geneva Observatory, University of Geneva \\
CH-1290 Sauverny, Switzerland \\
andre.maeder@unige.ch}



\begin{abstract}
The hypothesis is made that, at large scales where General 
Relativity may be applied, the empty space  is  scale invariant.
 This establishes a relation between the  cosmological constant   and  
  the scale factor  $\lambda$ of the scale invariant framework.
This relation  brings major simplifications in the scale invariant
 equations for cosmology, which contain a new term, depending 
on the derivative of the scale factor,  that  opposes to gravity and produces an accelerated expansion.
The displacements due to the acceleration term  make a high contribution
 $\Omega_{\lambda} $ to the energy-density of the Universe, satisfying an equation 
of the form $ \; \Omega_{\mathrm{m}}+\Omega_{\mathrm{k}}+\Omega_\lambda =1 $.
The models do not demand the existence of unknown particles.
 There is a  family of flat models 
with different density parameters $\Omega_{\mathrm{m}} < 1$.

Numerical integrations of the cosmological equations for different values of the curvature and
  density parameter $k$ and $\Omega_{\mathrm{m}}$ are performed. 
  The presence of even tiny amounts of matter in the Universe tends  to   kill scale invariance. The point 
  is that for $\Omega_{\mathrm{m}}$= 0.3  the effect is not yet completely killed.
 The models with non-zero density
 start explosively with first a braking phase followed by a continuously  accelerating expansion.
 Several observational properties are examined, in particular the distances, the m--z diagram,
 the $\Omega_{\mathrm{m}}$ vs. $\Omega_{\lambda}$ plot.
Comparisons with observations are also performed for  the Hubble constant 
 $H_0$ vs.  $\Omega_{\mathrm{m}}$,
for the expansion history 
in the plot $H(z)/(z+1)$ vs. redshift $z$ and for the transition redshift from braking to acceleration.
 These first dynamical tests are satisfied by the scale invariant models,
 which thus deserve further studies.
 
\end{abstract}




\section{Introduction}  \label{sec:intro}
 
The cause  of the accelerating expansion of the Universe \citep{Riess98,Perl99} is a
major scientific problem at present,  see for example the reviews by
\citet{Wein89},  \citet{Peeb03}, \citet{Frie08}, \citet{Feng10}, \citet{Port11} and  \citet{Sola13}. Here, we  explore whether scale invariance 
may bring something interesting in this context.

The laws of physics   are generally  not unchanged under a change of scale,
 a fact  discovered by Galileo Galilei  
\citep{Feynman63}. The scale references 
are closely related  to the material content of the medium.  Even the vacuum
at the quantum level is  not scale invariant, since quantum effects produce some units
of time and length. As a matter of fact, this  point  is a historical argument,
 already forwarded long time ago against  Weyl's  theory \citep{Weyl23}.

 The  empty space at  macroscopic and  large scales
does not  have any  preferred scale of length or
time.  The key--hypothesis of this work,  is that the empty space at  large scales  is scale invariant. 
The astronomical scales 
 differ by an enormous factor from the nuclear scales where quantum effects intervene.
Thus, in the same way as we may use the  Einstein theory at large scales, 
even if we do not have a quantum theory of gravitation,
we may consider that the large scale empty space is scale invariant even if this  is not true at   the  quantum level.
Indeed, the results of the models below will show that the possible cosmological effects  of scale invariance tend to  very rapidly
 disappear when some matter is present in the Universe (cf. Fig.~\ref{scale}).
The aim of this paper is to  carefully examine the implications of the hypothesis of scale invariance in cosmology,
  accounting in particular for the  specific hypothesis of  scale invariance of the empty space at large. 
This  is in the line of the remark by \citet{Dirac73} that the  equations
expressing the basic laws of Physics should be invariant under the widest possible group of transformations. This is the case of the 
Maxwell equations of electrodynamics which  in absence of charges and currents show the property of scale invariance.
While scale invariance has often been studied in relation with possible variations of the gravitational constant $G$,
we emphasize that no such hypothesis  is considered here.

In Sect. 2, we briefly  recall  the basic scale invariant field equations necessary 
for the present study and apply them  to the empty space at  large scales. 
In Sect. 3, we establish the basic equations of the scale invariant cosmology accounting
for the properties of the empty space.
We also examine  the density and geometrical parameters.
In Sect. 4, we present  numerical results of scale invariant models for different
curvature parameters $k$  and density parameters, and we make some comparisons with $\Lambda$CDM models.
In Sect. 5, we examine several basic observational properties of the models
and perform some first comparisons with model independent observations. 
 Sect. 6 briefly gives the conclusions.

\section{The hypothesis of  scale invariance of the empty space and its consequences}   \label{sect:empty}

\subsection{Scale invariant field equations}  \label{sub:field}

 Many  developments on a scale invariant theory of gravitation have been performed 
by physicists like \citet{Weyl23},  \citet{Eddi23},  \citet{Dirac73}, and \citet{Canu77}. 
Here, we limit the recalls to the very minimum. In the
 4-dimensional space of General Relativity (GR), the element interval  
 $ds' \,^2$  writes $ ds'\,^2  \,= \, g'_{\mu \nu} dx\,^{\mu} \, dx\,^{\nu}$,
(symbols with a prime refer to the space of GR, which is not scale invariant).
A scale (or gauge) transformation is a change  of the line element, for example  $ds'$,  of the form,   
\begin{equation}
ds' \, = \, \lambda(x^{\mu}) \, ds \, .
\label{lambda}
\end{equation}
\noindent
There,   $ds^2  =  g_{\mu \nu} dx^{\mu} \, dx^{\nu} $
is the line element in another  framework, where  we now consider scale invariance as a fundamental property in addition 
to  general covariance, (the coordinate intervals are dimensionless).
Parameter $\lambda(x^{\mu})$ is the scale factor connecting the two line elements.
 There is a conformal transformation between the two systems,
\begin{equation}
 g'_{\mu \nu} \, = \, \lambda^2 \, g_{\mu \nu} \, .
 \label{conformal}
 \end{equation}
 \noindent
 The Cosmological Principle   demands
that the scale factor   only  depends on time.
Scalars, vectors or
tensors  that  transform like
\begin{equation}
Y'^{\, \nu   }_{\mu}  \, =  \, \lambda^{n} \, Y^{\nu}_{\mu} \, ,
\label{co}
\end{equation}
\noindent
 are respectively called coscalars, covectors or  cotensors  of power $n$. 
 If $n=0$, one has an inscalar, invector or intensor, such objects are
 invariant to a scale transformation  like  (\ref{lambda}). Scale covariance  refers to a transformation (\ref{co}) 
 with  powers $n$ different from zero, while we reserve the term scale invariance  for cases with    $n=0$.
An extensive cotensor analysis has been developed by the above mentioned authors, see also \citet{Bouv78}.  The derivative
 of a  scale invariant object is not in general scale invariant.
Thus,  scale covariant derivatives of the first and second order have been developed preserving  scale covariance.
 Modified Christoffel symbols,  Riemann-Christoffel tensor  $R^{\nu}_{\mu \lambda \rho}$,  
 its contracted form  $R^{\nu}_{\mu}$ and the total
curvature $R$ also have their scale invariant 
corresponding terms  which were developed and studied by \citet{Weyl23}, \citet{Eddi23}, \citet{Dirac73} and \citet{Canu77}. 
The last reference provides a summary of scale invariant tensor analysis.
The main difference with standard tensor analysis is that all these ''new''  expressions contain  terms depending  on 
$\kappa_{\nu}$, which is a derivative of the above  scale factor $\lambda$,
\begin{equation}
\kappa_{\nu} \, = \, -\frac{\partial}{\partial x^{\nu}  } \, \ln \lambda \, .
\label{kappa}
\end{equation}
\noindent
This term is called the coefficient of metrical connection.
It is as a fundamental quantity as are the $g_{\mu \nu}$ in GR. 
 We jump straightforward to the expressions of the contracted form $R^{\nu}_{\mu}$ of the
  Riemann-Christoffel, derived by \citet{Eddi23} and \citet{Dirac73} in the scale invariant system,
\begin{equation}
R^{\nu}_{\mu} = R'^{\nu}_{\mu}  - \kappa^{; \nu}_{\mu}  - \kappa^{ \nu}_{;\mu}
 - g^{\nu}_{\mu}\kappa^{ \alpha}_{;\alpha}  -2 \kappa_{\mu} \kappa^{\nu}
+ 2 g^{\nu}_{\mu}\kappa^{ \alpha} \kappa_{ \alpha}  \, .
\label{RC}
\end{equation}
There,  $R'^{\nu}_{\mu}$ is the usual expression in the Riemann geometry. We see that the additional terms with respect to GR are
 all only depending on $\kappa_\nu$ (which is also the case for the general field equation below).
A sign '';'' indicates a derivative with respect to the mentioned coordinate. The total curvature $R$ in the scale invariant system  is 
\begin{equation}
R \, = \, R' -6 \kappa^{\alpha}_{; \alpha}+6 \kappa^{\alpha} \kappa_{\alpha} \, .
\label{RRR}
\end{equation}
\noindent
There,   $R'$ is the total curvature in Riemann geometry
The above expressions allow one to write the  general scale invariant field equation \citep {Canu77},
\begin{eqnarray}
R'_{\mu \nu}   -  \frac{1}{2}  \ g_{\mu \nu} R' - \kappa_{\mu ;\nu}  - \kappa_{ \nu ;\mu}
 -2 \kappa_{\mu} \kappa_ {\nu}   
+ 2 g_{\mu \nu} \kappa^{ \alpha}_{;\alpha}
 - g_{\mu \nu}\kappa^{ \alpha} \kappa_{ \alpha}   =
-8 \pi G T_{\mu \nu} - \lambda^2 \Lambda_{\mathrm{E}}  \, g_{\mu \nu} \, ,
\label{field}
\end{eqnarray}
\noindent
where $G$ is the gravitational constant (taken here as a true constant) and
 $\Lambda_{\mathrm{E}}$ the Einstein cosmological constant.
The  second member of (\ref{field}) is scale invariant,  as is the first one.
This means that
\begin{equation}
T_{\mu \nu} \, =  \,T '_{\mu \nu} \, .
\end{equation}
\noindent
This has further implications,  as illustrated in the case of a perfect fluid by \citet{Canu77}. 
The tensor  $T_{\mu \nu}$  being  scale invariant,  one may write
$
 ( p+\varrho) u_{\mu} u_{\nu} -g_{\mu \nu } p =
   ( p'+\varrho') u'_{\mu} u'_{\nu} -g'_{\mu \nu } p' \, .
$
The velocities $u'^{\mu}$ and $u'_{\mu}$ transform like
\begin{eqnarray}
u'^{\mu}=\frac{dx^{\mu}}{ds'}=\lambda^{-1}  \frac{dx^{\mu}}{ds}=  \lambda^{-1} u^{\mu} \, ,  \nonumber \\
  \; \mathrm{and} \; \;
u'_{\mu}=g'_{\mu \nu} u'^{\nu}=\lambda^2 g_{\mu \nu} \lambda^{-1} u^{\nu} = \lambda \, u_{\mu} \, .
\label{pl1}
\end{eqnarray}
\noindent
Thus, the energy-momentum tensor transforms like 
\begin{equation}
( p+\varrho) u_{\mu} u_{\nu} -g_{\mu \nu } p  =
 ( p'+\varrho') \lambda^2 u_{\mu} u_{\nu} - \lambda^2 g_{\mu \nu } p' \, .
 \label{pl}
\end{equation}
\noindent
This implies the following scaling of $p$ and $\varrho$ in the new general coordinate system
\begin{equation}
 p =  p'  \, \lambda^2    \, \quad \mathrm{and} \quad   \varrho =  \varrho'  \, \lambda^2 \, .
\label{ro2}
\end{equation}
\noindent
Pressure and density are therefore not scale invariant, but are so-called coscalars of power  -2. 
 The term containing the cosmological
constant now writes  $ -\lambda^2 \Lambda_{\mathrm{E}}  \, g_{\mu \nu}$, it is thus also a cotensor of power -2.
To avoid any ambiguity, we  keep all  expressions with $\Lambda_{\mathrm{E}}$,
the  Einstein cosmological constant,
so that in the basic scale invariant field equation (\ref{field}), it appears multiplied by $\lambda^2$. 

\subsection{Application to the empty space}   \label{sub:appl}
We now consider the application of the above scale invariant equations to the empty  space.  
In GR, the corresponding $ g'_{\mu \nu}$ represent the de Sitter metric
 for an empty  space with  $\Lambda_{\mathrm{E}}$ different from zero. The de Sitter metric can be written in various ways. 
 An  interesting form  originally found by Lemaitre
and by Robertson \citep{Tolman34} is 
\begin{equation}
ds'^2 = dt^2- e^{2kt}[dx^2 + dy^2 +dz^2] \, ,
\end{equation}
 \noindent
 with the prime referring to the system of GR.  In this form, the $g'_{\mu \nu}$ 
 are not independent of the time coordinate.
 Also, we have $k^2 = \Lambda_{\mathrm{E}}/3 $ and $c=1$.
 Now, a  transformation of the coordinates can be made  \citep{Mavrides73},
 \begin{equation}
 \tau \, = \, \int  e^{\left(- \sqrt{\frac{\Lambda_{\mathrm{E}}}{3} } \, t  \right)}   dt \, ,
 \label{int}
\end{equation} 
\noindent
where $\tau$ is a new time coordinate.
This allows one to show that the de  Sitter metric is conformal  to 
 to the Minkowski metric. With the above transformation, the de Sitter line element $ds'^2$ may be written
 \begin{equation}
 ds'^2 \, = \, e^{\psi (\tau)} [d\tau^2 -(dx^2 + dy^2 +dz^2)] \, , \quad \mathrm{with} \quad 
  e^{\psi (\tau)} \, =  \,  e^{\left( 2 \, \sqrt{\frac{\Lambda_{\mathrm{E}}}{3} } \, t  \right)} \, .
  \label{eps}
 \end{equation}
 \noindent
  From  this and relation (\ref{int}), we also have
   \begin{equation}
  e^{\psi (\tau)} \, =    \, \frac{3}{\Lambda _{\mathrm{E}}\, \tau^2 } \, .
  \end{equation}
  \noindent
  Thus,   the corresponding  line element in the scale invariant system can be written
 \begin{equation}
 ds^2 \, = \, \lambda^{-2} ds '^2 \, = \, \frac{3 \, \lambda^{-2} }{\Lambda  _{\mathrm{E}}\, 
  \tau ^2 } \, [c^2 \, d\tau^2 -(dx^2 + dy^2+dz^2)] \, .
\end{equation} 
 \noindent
 Thus, the de Sitter metric is conformal to the Minkowski metric. Furthermore, if
 the following relation happens to be valid,
 \begin{equation}
  \frac{3 \, \lambda^{-2} }{\Lambda  _{\mathrm{E}}\,   \tau^2 } \, = \, 1  \, ,
  \label{equal}
  \end{equation}
 \noindent
 then the  line-element  for the empty scale invariant space   would just be  given by the Minkowski metric,
 \begin{equation}
ds^2 \, = \,  d\tau^2 - (dx^2+dy^2+dz^2) \,  .
\label{Mink}
\end{equation}
\noindent
Now, we take as a hypothesis that the Minkowski metric  (\ref{Mink}) which characterizes Special Relativity 
also applies in the scale invariant framework.
However, we  have  to check whether this hypothesis is consistent, {\it{i.e}} whether
the Minkowski metric is a choice permitted by  the scale invariant field equation (\ref{field}),  and also
 whether  relation (\ref{equal}) is consistent.
The Minkowski metric implies that in  (\ref{field}) one has,
\begin{equation} 
R'_{\mu \nu}  \, - \, \frac{1}{2}  \ g_{\mu \nu} \, R'  \, = \, 0 \, \quad \mathrm{and} \quad T'_{\mu \nu} =0 \,  .
\end{equation}
\noindent
Thus,   only  the following terms are remaining from  the above  equation (\ref{field}), 
\begin{equation}
  \kappa_{\mu ;\nu}  + \kappa_{ \nu ;\mu}
 +2 \kappa_{\mu} \kappa_{\nu}
- 2 g_{\mu \nu} \kappa^{ \alpha}_{;\alpha}
 + g_{\mu \nu}\kappa^{ \alpha} \kappa_{ \alpha}   =  \lambda^2 \Lambda_{\mathrm{E}}  \, g_{\mu \nu}  .
\label{fcourt}
\end{equation}
\noindent
We are left  with a relation  between some functions of the scale factor $\lambda$
(through the $\kappa$-terms (\ref{kappa})), the $g_{\mu \nu}$  and the Einstein cosmological constant 
$\Lambda_{\mathrm{E}}$,   which now  appears as related to the properties of scale invariance of the empty space.
Since $\lambda$ may only be a function of time (which we now call $t$ instead of $\tau$),
only the zero component
of $\kappa_\mu$ is non-vanishing. Thus, the coefficient of metrical connection becomes
\begin{eqnarray}
 \kappa_{\mu ;\nu} = \kappa_{0 ;0} = \partial_0 \kappa_0 =  \frac{d \kappa_0}{dt} \equiv
{\dot{\kappa}_0} \, .
\end{eqnarray}
\noindent
The 0 and   the 1, 2, 3 components  of what remains from the field equation (\ref{fcourt})  become respectively
\begin{equation}
3 \kappa^2_0 \, = \,\lambda^2 \, \Lambda_{\mathrm{E}} \, ,
\label{k1}
\end{equation}
\begin{equation}
 2  \dot{\kappa}_0 - \kappa_0^2 = -\lambda^2  \Lambda_{\mathrm{E}}   \, .
\label{k2}
\end{equation}
\noindent
From  (\ref{kappa}), one has $\kappa_0 \, = - \dot{\lambda}/\lambda$ (with  $c=1$ at the denominator),
thus  (\ref{k1}) and  (\ref{k2}) lead to
\begin{eqnarray}
\  3 \, \frac{ \dot{\lambda}^2}{\lambda^2} \, =\, \lambda^2 \,\Lambda_{\mathrm{E}}  \,  
 \quad \mathrm{and} \quad 2 \, \frac{\ddot{\lambda}}{\lambda} - \frac{ \dot{\lambda}^2}{\lambda^2} \, =
\, \lambda^2 \,\Lambda_{\mathrm{E}}  \, .
\label{diff1}
\end{eqnarray}
\noindent
These expressions   may also be written in equivalent forms
\begin{eqnarray}
\frac{\ddot{\lambda}}{\lambda} \, = \,  2 \, \frac{ \dot{\lambda}^2}{\lambda^2} \, , \quad
 \quad \mathrm{and} \quad \frac{\ddot{\lambda}}{\lambda} -\frac{ \dot{\lambda}^2}{\lambda^2} \,
  = \, \frac{\lambda^2 \,\Lambda_{\mathrm{E}}}{3} \, .
\label{diff2}
\end{eqnarray}
\noindent
The first relation of (\ref{diff2}) places a constraint on $\lambda(t)$.
Considering  a solution of the form
$\lambda \,= \, a\,  (t \,- \,b)^n+d $, we get
$d \,= \,0$  and  $n \, = \, -1$. There is no
condition on $b$, which means that the  origin $b$ of the timescale is not determined by the above equations.
 (The origin of the time will be determined by 
 the solutions of the equations of the particular cosmological model considered, see Sect. 4).
With the first of equations  (\ref{diff1}), we get
\begin{equation}
\lambda \, = \, \sqrt{\frac{3}{\Lambda_{\mathrm{E}}}} \, \frac{1}{c \,t}  \, .
\label{lamb}
\end{equation}
\noindent
For numerical estimates,  $c$ is indicated here. This condition is the same as (\ref{equal}) obtained from the above conformal transformation.
We have thus verified that the Minkowski metric is compatible
with the scale invariant field equation  for the above condition.  

The problem of the cosmological constant in the empty space is not a new one.
First, in the context  of RG,  models  with a  non-zero cosmological constant, such as  the $\Lambda$CDM models, do not admit the
Minkowski metric, which may be a problem. Here, one can also consistently apply Special Relativity (with the Minkowski metric), 
even if the cosmological constant $\Lambda_{\mathrm{E}}$ is not equal to zero.
Also \citet{Bert90} are  quoting  the following remark
they got from   Bondi: {\it{ ``Einstein's disenchantment with the cosmological 
constant was partially motivated by a desire to preserve scale invariance
of the empty space Einstein equations ".}} This remark is  consistent  with the fact
 that $\Lambda_{\mathrm{E}}$ is not scale invariant 
as shown by the fact that $\Lambda_{\mathrm{E}}$ is multiplied by $\lambda^2$ in (\ref{field}). 
  Thus, the   scale invariant framework  offers a possibility
to conciliate  the existence of $\Lambda_{\mathrm{E}}$ with the scale invariance of the empty space.


\section{Cosmological  equations and their properties} 
 
 \subsection{Basic scale invariant cosmological equations} \label{sub:cosmequ}
The metric appropriate  to cosmological models is the Robertson-Walker metric, characteristic of the homogeneous and isotropic
space. We need to obtain the differential equations constraining  the expansion factor $R(t)$ in the scale invariant framework.
These equations can be derived in various equivalent ways  \citep{Canu77}: -- by  expressing
the  general cotensorial field equation (\ref{field}) with the Robertson-Walker metric, 
 -- by taking advantage that there is a conformal transformation
between the metrics $g'_{\mu \nu}$ (GR) and $g_{\mu \nu}$ (scale invariant), -- and  by applying a scale transformation 
 to the current differential  equations of cosmologies in $R(t)$. 
The scale invariant  equations are,
\begin{equation}
\frac{8 \, \pi G \varrho }{3} = \frac{k}{R^2}+
\frac{\dot{R}^2}{R^2}+ 2 \, \frac{\dot{\lambda} \, \dot{R}}{\lambda \, R}+
\frac{\dot{\lambda}^2}{\lambda^2} - \frac {\Lambda_{\mathrm{E}} \lambda^2}{3} \,
\label{E1p}
\end{equation}
\begin{equation}
-8 \, \pi G p = \frac{k}{R^2}+ 2 \frac{\ddot{R}}{R} + 2 \frac{\ddot{\lambda}}{\lambda}+\frac{\dot{R}}{R}^2
+ 4 \frac{\dot{R} \, \dot{\lambda}}{R \, \lambda}-\frac{\dot{\lambda^2}}{\lambda^2} -\Lambda_{\mathrm{E}} \,  \lambda^2  \, .
\label{E2p}
\end{equation}
\noindent
These equations contain several additional terms with respect to the standard case.
We now also account for expressions (\ref{diff1}) and (\ref{diff2}) for the empty space, which  characterize $\lambda$ 
 and its properties. 
 This is consistent with ''the postulate of GR that gravitation couples universally
 to all energy and momentum"  \citep{Carr92}. This also applies to the energy associated to  the derivatives of 
 $\lambda$    in Eq. (\ref{E1p}) and (\ref{E2p}),  which are responsible  for an accelerated
 expansion as we will see below.
 Thus, with   (\ref{diff1}) and (\ref{diff2}), the two above cosmological equations (\ref{E1p}) and
(\ref{E2p})  nicely simplify to
\begin{equation}
\frac{8 \, \pi G \varrho }{3} = \frac{k}{R^2}+\frac{\dot{R}^2}{R^2}+ 2 \,\frac{\dot{R} \dot{\lambda}}{R \lambda} \, ,
\label{E1}
\end{equation} 
\begin{equation}
-8 \, \pi G p  = \frac{k}{R^2}+ 2 \frac{\ddot{R}}{R}+\frac{\dot{R^2}}{R^2}
+ 4 \frac{\dot{R} \dot{\lambda}}{R \lambda}  \, .
\label{E2}
\end{equation}
\noindent
The combination of these two equations  leads to
\begin{equation}
-\frac{4 \, \pi G}{3} \, (3p +\varrho)  =  \frac{\ddot{R}}{R} + \frac{\dot{R} \dot{\lambda}}{R \lambda}  \, .
\label{E3}
\end{equation}
\noindent
Term $k$ is the curvature parameter which  takes values $0$ or $\pm 1$,  
$p$ and $\varrho$ are the pressure and density in the scale invariant system.
 Einstein cosmological constant has disappeared due to the account of the properties of the empty space at   large scales.
These three equations differ from the classical ones, in each case  only by an  additional 
term containing  $\frac{\dot{R} \dot{\lambda}}{R \lambda} $, which depends on time $t$.
If $\lambda(t)$ is a constant, 
one gets the usual equations of cosmologies. 
Thus at any fixed time, the effects that do not depend on time evolution are just those 
predicted by GR  (for example, the gravitational shift in stellar spectral lines).
Significant departures from  GR may
 appear in cosmological evolution over the ages. 

  What is the significance of  the  additional term?
Let us consider (\ref{E3}):
the term $\frac{\dot{R} \dot{\lambda}}{R \lambda}$ is negative, since  according to
(\ref{lamb}) we have 
$\dot{\lambda}/\lambda \, = \, - \frac{1}{t}$. 
This term  
  represents  {\emph{ an acceleration that opposes the gravitation}}.
The term   $\frac{\dot{R} \dot{\lambda}}{R \lambda}$ is equal to  $ - \frac{H}{t} $.  
Thus, 
equations  (\ref{E1}) to (\ref{E3}) are fundamentally different from those  of the $\Lambda$CDM models: a variable term replaces the 
cosmological constant $\Lambda_{\mathrm{E}}$. The new term  implies that the acceleration of the expansion  is variable in time,
a suggestion which is not so new \citep{Peeb88}.
 In this connexion, we note that the need to have  a time-dependent  term
in order to satisfactorily interpret the recent observations has been recently emphasized by several authors,
\citet{Sahni14}, \citet{Sola15}, \citet{Ding15} and \citet{Sola16}.

\subsection{The case of an empty Universe} \label{sub:empty}
 Empty Universe models (like the de Sitter model) are interesting as they represent the asymptotic limit
of models with lower and lower densities. 
Let us  consider  the   case of a  non-static scale invariant  model  with no matter 
and no radiation ($\varrho$ = 0  and $p$ = 0). The corresponding  
Friedman model  would have  an expansion given by $R(t) \sim t$ (with $k=-1$). 
In our case, expression (\ref{E3}) becomes simply
\begin{equation} 
\frac{\ddot{R}}{R} \, = \, \frac{\dot{R} \dot{\lambda}}{R \lambda}  \,  .
\label{E3vide}
\end{equation}
\noindent
The integration   gives $\dot{R}=a \,t$, where $a$ is a constant, a further integration gives
$R = a(t^2-t^2_{\mathrm{in}})$.
The initial instant $t_{\mathrm{in}}$   is  chosen at the origin  $R(t_{\mathrm{in}})=0$  of the 
considered model. The value of  $t_{\mathrm{in}}$ is not necessarily 0 and we do not know it yet. The model is non-static and 
the Hubble value  at time $t$ is 
\begin{equation}
H = \frac{\dot{R}}{R} = 2  \frac{t}{(t^2-t^2_{\mathrm{in}})} \,  
\label{H}
\end{equation}
\noindent
To get more information,  we  use  Eq. (\ref{E1}) and get,
\begin{equation}
\dot{R}^2 \, t - 2 \, \dot{R}  R + k t =0   \, ,
\end{equation}
\noindent
which   leads to a second  expression for $H$,
$H=\frac{\dot{R}}{R}= \frac{1}{t} \pm \frac{\sqrt{1- k\frac{t^2}{R^2}}}{t}.$
For an empty model, we may take $k=-1$  or $k=0$ (a choice consistent with (\ref{kom}) 
 in the study of the geometrical parameters below).
Let us first consider the case $k=-1$, (the dimensions of $k$ go like $[R^2/t^2]$).
 Fixing the scale so that  $t_0 =1$ and $R_0=1$ at the present time, we get ($H_0$ being positive),
\begin{equation}
H_0= {1+\sqrt{2}}\,.     
\label {Hovide}
\end{equation}
\noindent
The above value  (for $k=-1$) represents a lowest bound  of $H_0$-values  to the  models with
non-zero densities, (since the steepness of  $R(t)$ increases with higher densities according to (\ref{E1})).
In (\ref{Hovide}), $H_0$ is   expressed as a function of $t_0=1$, however the  value of $H_0$ should rather  
be expressed in term of the age  $\tau= t_0 -t_{\mathrm{in}}$ of the Universe in the considered model. 
We find $ t_{\mathrm{in}}$ by expressing the equality of the two values of $H_0$ obtained above,
\begin{equation}
\frac{t_{\mathrm{in}}}{t_0} \,= \,  \sqrt{2} - 1 \, .
\end{equation}
\noindent 
This is  the minimum value of $t_{\mathrm{in}}$, (we notice that the scale factor $\lambda (t)$ at $t_{\mathrm{in}}$ has a  value limited 
by $1+\sqrt{2}=2.4142$).
The corresponding age $\tau$ becomes $\tau = (2-\sqrt{2}) \, t_0$  and  we may now 
 express $H_0$ from (\ref{Hovide}) as a function of the  age $\tau$.
We have quite generally, indicating here in parenthesis the timescale referred to,
\begin{equation}
\frac{H_0(\tau)}{\tau} \, = \, \frac{H_0(t_0)}{t_0} \, , \quad  \mathrm{thus} \; H_0(\tau) \, =   \, H(t_0) \, \tau  \, .
\label{htau}
\end{equation}
\noindent
Thus, we get the following value of $H_0 (\tau)$  (which appears as  a maximum), 
\begin{equation}
H_0 (\tau)\,  = \, \sqrt{2} \,  \quad \mathrm{for} \; \; k= -1 \, .
\label{hovide}
\end{equation} 

Let us now turn to the  empty model with  $k=0$. We  have  from what precedes
$H_0 \, = \, {2} $.
 As for $k=-1$, this value for the empty space is the minimum value of $H_0$ expressed
in the scale with $t_0=1$.
The comparison  of this value and ({\ref{H}) leads
to $t_{\mathrm{in}}=0$ and thus $ \tau=t_0$. Here also, $t_{\mathrm{in}}=0$ is the minimum value for all  models with $k=0$
and thus  $\tau$ is the longest  age.  We see that for $k=0$, the scale factor $\lambda$ of the empty space  tends towards  infinity at the origin.
Expressing $H_0$ as a function of $\tau$, we have
$H_0(\tau) \, = \, 2.$
Here also, $H_0(\tau)$ is an upper bound  for the models with $k=0$. On the whole, empty models, 
whether $k=-1$ or $k=0$, obey very simple properties.

The empty  scale invariant model    expands  like $t^2$,   faster than the linear  corresponding Friedman model. 
The effects of scale invariance appear as the source
of an accelerated expansion, consistently with the remark made  about  (\ref{E3}).
  
\subsection{Critical density and $\Omega$-parameters}  \label{sub:density}

We  examine some general properties of the models
based on equations (\ref{E1})  to (\ref{E3}).
 The critical density  corresponding to 
 the flat space with  $k=0$ is from (\ref{E1}) and $\lambda =1/t$, with the Hubble parameter $H=\dot{R}/R$,
\begin{equation}
\frac{8 \, \pi G \varrho^{*}_{\mathrm{c}} }{3} \, =\, H^2 -2\, \frac{H}{t}  \, .
\label{rocm1}
\end{equation}
\noindent
 We mark with a  '' * ''  this critical density that is different from the usual definition. The second member is always $\geq 0$, 
 since 
 the relative  growth rate  for non empty models  is  higher than $t^2$.
With (\ref{E1}) and (\ref{rocm1}), we have
\begin{equation}
\frac{\varrho}{\varrho^*_{\mathrm{c }}} - \frac{k}{R^2 H^2} + \frac{2}{H t}
\left(1 - \frac{\varrho}{\varrho^*_{\mathrm{c }}} \right)= \, 1 \, .
\label{sum}
\end{equation}
\begin{equation}
\mathrm{With}  \;  \; \; \Omega*_{\mathrm{m}} = \frac{\varrho}{\varrho*_{\mathrm{c }}} , \, \; \mathrm{and} \, \;
\Omega_{\mathrm{k}} = -\frac{k}{R^2  H^2} \, ,
\label{Ok}
\end{equation}
\noindent
we get
\begin{equation}
\Omega^*_{\mathrm{m}}  +  \Omega_{\mathrm{k}} + \frac{2}{H \, t} \left( 1 - \Omega^*_m \right)  =  1  .
\label{Omega}
\end{equation}
\noindent
Consistently, $\Omega^*_{\mathrm{m}}=1$     implies $\Omega_{\mathrm{k}}=0$ and reciprocally.
We also consider the usual definition of the critical density,
\begin{equation}
\Omega_{\mathrm{m}}= \frac{\varrho}{\varrho_{\mathrm{c }}}\, \quad  
\mathrm{with} \quad  \varrho_{\mathrm{c }}= \frac{ 3 \, H^2}{8 \pi G} \, .
\label{rhocc}
\end{equation}
\noindent
Between the two density parameters we have the relation
\begin{equation}
\Omega_{\mathrm{m}} = \Omega^*_{\mathrm{m}} \left( 1 - \frac{2}{H t} \right) \, .
\label{relom}
\end {equation}
\noindent
From (\ref{rocm1}) or (\ref{Omega}),    we obtain   
\begin {equation}
\Omega_{\mathrm{m}} \, + \, \Omega_{\mathrm{k}} \, +  \Omega_{\lambda} = \, 1  \, ,
\quad \mathrm{with} \quad  \quad\Omega_{\lambda}  \equiv  \frac{2}{ H \, t} \, .
\label{Omegapr}
\end{equation}
\noindent
The  above relations  are  generally considered at time $t_0$, but they are also valid at other epochs.
  The term 
 $\Omega_{\lambda}$ arising from scale invariance has replaced the usual term $\Omega_\Lambda$.
The difference of the physical meaning is very  profound. 
While  $\Omega_\Lambda$ is the density parameter associated to  the cosmological constant or to the dark energy,
 {\emph{ $\Omega_\lambda$ expresses the energy density resulting from the variations of the scale factor.}}
This  term does not demand the existence of  unknown particles.  It is interesting  that  an  equation  like  (\ref{Omegapr}) 
is also valid in the scale invariant cosmology. 
%

While in Friedman's models there is only one model corresponding to $k=0$, 
here for  $k=0$ there is 
    a variety of models with different  $\Omega_{\mathrm{m}}$  and $\Omega_{\lambda}$. 
For all models, whatever the $k$-value,  the density parameter
  $\Omega_{\mathrm{m}}$ remains smaller than 1,  (this does not apply to $\Omega^*_{\mathrm{m}}$  for $k=+1$). 
  For $k=-1$ and $k=0$, this is clear   since  $2 \, t_0/H_0$ is always positive.
  For $k=+1$ (negative  $\Omega_{\mathrm{k}}$), numerical models   confirm that $  \Omega_{\mathrm{k}} \, + 
   \Omega_{\lambda} $ is always positive so that $\Omega_{\mathrm{m}} < 1$. Thus, a density parameter $\Omega_{\mathrm{m}}$
always smaller than 1.0 appears as a fundamental   property of scale invariant models.

\subsection{The geometry parameters}  \label{sub:geometry}

We now consider  the geometry parameters $k$, $q_0 = - \frac{\ddot{R}_0 \, R_0}{\dot{R}^2_0}$
 and their relations with $\Omega_{\mathrm{m}}$, $\Omega_{\mathrm{k}}$ and $\Omega_{\lambda}$ at the present time $t_0$.
Expression (\ref{E2}), in absence of pressure and 
divided by $H^2_0$,  becomes
\begin{equation}
-2 \, q_0 +1 - \Omega_{\mathrm{k}}= \frac{4}{H_0 t_0} \, .
\label{qk}
\end{equation}
\noindent
Eliminating $\Omega_{\mathrm{k}}$ between (\ref{qk}) and (\ref{Omega}), we   obtain
\begin{equation}
2 \, q_0 =  \Omega^*_{\mathrm{m}}- \frac{2}{H_0 t_0} ( \Omega^*_{\mathrm{m}}+1)  \, ,
\label{int1}
\end{equation}
\noindent
and  thus, using $\Omega_{\mathrm{m}}$ rather  than $\Omega^*_{\mathrm{m}}$, we get}
\begin{equation}
q_0 \,= \, \frac{\Omega_{\mathrm{m}}}{2}-  \frac{\Omega_{\mathrm{\lambda}}}{2} \, .
\label{qzerom}
\end{equation}
This  establishes a relation between the deceleration parameter $q_0$ and the matter content for  the
scale invariant cosmology. If $k=0$, we have  
\begin{equation}
q_{0} = \frac{1}{2} -\Omega_{\lambda} \, =  \, \Omega_{\mathrm{m}} - \frac{1}{2} \, ,
\label{qkk}
\end{equation}
which provides a very simple relation between basic parameters. For  $\Omega_{\mathrm{m}}=0.30$ or 0.20, we would
get $q_0=-0.20$ or -0.30.
The above basic relations are different from those of the $\Lambda$CDM, which  is  expected 
since the basic equations (\ref{E1}) - (\ref{E3}) are different. Let us recall
that in the $\Lambda$CDM model with $k=0$ one has
$q_0\, = \, \frac{1}{2}\Omega_{\mathrm{m}} -\Omega_{\Lambda}$.
 For $\Omega_{\mathrm{m}}=0.30$ or 0.20, we get  $q_0=-0.55$ or -0.70.
In both cosmological models, one has  simple, but different,  relations expressing 
the $q_0$ parameter.

Let us now turn to the curvature parameter $k$. From the basic equation (\ref{E1})
and the definition of the critical density (\ref{rocm1}), this becomes,
\begin{equation}
\frac{k}{R^2_0} =H^2 _0\left[ (\Omega^*_{\mathrm{m}} -1) \left(1 - \frac{2}{t_0 H_0} \right )\right]  \, , 
\label{kom}
\end{equation}
\noindent
 which establishes a relation between $k$ and $\Omega^*_{\mathrm{m}}$.  It confirms that
 if $\Omega^*_{\mathrm{m}}=1 $, one also has $k=0$
and reciprocally. 
Values of $\Omega^*_{\mathrm{m}} > 1$ correspond to a positive $k$-value,
 values smaller than $1$ to a negative $k$-value.  With $\Omega_{\mathrm{m}}$, one has
\begin{equation}
\frac{k}{R^2_0} =H^2 _0\left[ \Omega_{\mathrm{m}}  -\left(1- \frac{2}{t_0 H_0} \right )\right]  \, , 
\label{kom2}
\end{equation}
\noindent
which is   equivalent to (\ref{Omegapr}) at time $t_0$. 
We also have
a relation between $k$ and $q_0$. From (\ref{int1})
and  (\ref{kom}), we obtain 
\begin{equation}
\frac{k}{R^2_0} = H^2_ 0 \left[ 2 \, q_0 -1 + \frac{4}{H_0 t_0} \right] \, .
\label{kq}
\end{equation}
\noindent 
For $k=0$, it gives the same relation as  (\ref{qkk}) above. We  emphasize that in all these expressions
$t_0$ is not the present age of the Universe, but just  the present time  in a scale where $t_0 = 1$. As in Sect. \ref{sub:empty},
the present age is  $\tau= t_0 -t_{\mathrm{in}}$, where  the initial time $t_{\mathrm{in}}$ depends on the considered model.\\

\subsection{Inflexion point in the expansion} \label{sub:inflex}

In the scale invariant cosmology, like in the  $\Lambda$CDM models, there are both a braking force due to gravitation and 
an acceleration force. There may thus be initial epochs  dominated by gravitational braking and other later epochs  by acceleration.
From  (\ref{qzerom}), valid at any epoch $t$, we see that  $q=0$ occurs  when
\begin{equation}
 \Omega_{\mathrm{m}} \, =  \, \Omega_{\lambda}  \, .
\label{oml}
\end{equation}
\noindent
 The gravitational term dominates
in the early epochs and the $\lambda$-acceleration dominates in more advanced stages. 
The higher the $\Omega_{\mathrm{m}}$-value,
the later the inflexion point occurs. The empty model (Sect. \ref{sub:empty}),
 where $R(t) \sim t^2$, is an exception showing only  acceleration.
For a flat model with $k=0$, we have at the inflexion point when
$\Omega_{\lambda} \, = \,  \Omega_{\mathrm{m}} = \frac{1}{2}  \, $,
the matter and the $\lambda$-contributions are equal and both equivalent to 1/2.
The inflexion points are identified for the  flat models  in Fig. \ref{Rtzero} and Table 1.
These results differ from those for the $\Lambda$CDM models.  We
have $q=0$   for a flat $\Lambda$CDM model  when
$ \frac{1}{2}  \,  \Omega_{\mathrm{m}} \, =  \, \Omega_{\Lambda}.$
 The acceleration term  needs only to reach the half
of the gravitational term to give $q=0$, while   in the scale invariant case the inflexion
point is reached for the equality of the two terms. 

\subsection{Conservation laws}

 An additional invariance in the equations necessarily influences 
  the  conservation laws.  Moreover, we have   accounted
for the scale invariance of the vacuum at macroscopic scales by using (\ref{diff1}) and  (\ref{diff2}). 
These hypotheses have an impact on the conservation laws.
We first rewrite  (\ref{E1}) as follows and take its derivative,
\begin{equation}
8 \, \pi G \varrho R^3 = 3\, k R+ 3 \, \dot{R}^2 R+ 6  \, \frac{\dot{\lambda}} {\lambda} \dot{R}  R^2 \, , 
\end{equation}
\begin{eqnarray}
\frac{d}{dt} (8 \, \pi G \varrho R^3 ) 
= -3 \dot{R} R^2  \,  
 \left[-\frac{k}{R^2}- \frac{\dot{R}^2}{R^2}-2 \frac{\ddot{R}}{R}  
-2  \frac{\ddot{R}}{ \dot{R}} \frac{\dot{\lambda}}{\lambda}-2  \frac{\ddot{\lambda}}{\lambda}
- 4 \frac{\dot{R} \dot{\lambda}}{R \lambda} +2 \frac{\dot{\lambda}^2}{\lambda^2}\right]  \, ,
\end{eqnarray}
\noindent
which contains many terms with $\lambda$ and its derivatives.
Eq. (\ref{E1}), (\ref{E3}) and (\ref{diff2}) lead to  many simplifications,
\begin{eqnarray}
\frac{d}{dt} (8 \, \pi G \varrho R^3 )  \,  
=-3\, \dot{R} R^2 \left[8 \, \pi G p+\frac{{R}}{ \dot{R}} \frac{\dot{\lambda}}{\lambda}
\left(8 \, \pi Gp+ \frac{8 \, \pi G \varrho}{3}\right) \right]  \, ,
\end{eqnarray}
\begin{equation}
3 \, \lambda \varrho dR + \lambda R d\varrho+ R \varrho d\lambda +3 \, p \lambda dR +3 p R d\lambda = 0 \, ,
\end {equation}
\begin{equation}
 \mathrm{and} \; \;  \; 3 \, \frac{dR}{R} + \frac{d \varrho}{\varrho}+\frac{d \lambda}{\lambda}+ 3 \, \frac{p}{\varrho}\; \frac{dR}{R}+
3 \, \frac{p}{\varrho} \; \frac{d\lambda}{\lambda} = 0 \, .
\label{conserv1}
\end{equation}
\noindent
This can also be written in a form rather similar to the usual conservation law,
\begin{equation}
\frac{d(\varrho R^3)}{dR} + 3 \, p R^2+ (\varrho+3\, p) \frac{R^3}{\lambda} \frac{d \lambda}{dR} = 0 \, .
\label{conserv2}
\end{equation}
\noindent
These last two equations express the law of  conservation of mass-energy  in the scale invariant cosmology.
 For a constant $\lambda$, we evidently recognize the usual 
  conservation law.
  We now write the equation of state in the general form,
$
P \, = \, w \,  \varrho $,  with $c^2$ =1,
where $w$ is  taken here as a constant.
The equation of conservation (\ref{conserv1}) becomes
$
 3 \, \frac{dR}{R} + \frac{d \varrho}{\varrho}+\frac{d \lambda}{\lambda}+ 3 \, w \,  \frac{dR}{R}+
3 \, w \, \frac{d\lambda}{\lambda} = 0$,
with the following simple integral,
\begin{equation}
\varrho \, R^{3(w+1)}  \,  \lambda ^{(3w+1)} \,= const. 
\label{3w}
\end{equation}
\noindent
For $w=0$, this is the case of ordinary matter of density
$\varrho_{\mathrm{m}}$ without pressure, 
\begin{equation}
\varrho_{\mathrm{m}} \, R^3  \, \lambda =const. 
\label{consm}
\end{equation}
\noindent
which means that the inertial  and gravitational mass (respecting the Equivalence principle) within a covolume should both     slowly  
increase over the ages. We do not expect any matter creation  as by \citet{Dirac73}
 and thus  the number of baryons should be a  constant.
  In GR, a certain curvature of  space-time is associated to a given mass. Relation (\ref{consm})
 means that the curvature of space-time associated to a mass is a slowly evolving function of the time in the Universe.
 For an
empty model with $k=0$, the change  of $\lambda$ would be enormous from $\infty$ at $t$=0 to 1 at $t_0$.  In a flat model with 
$\Omega_{\mathrm{m}}=0.30$, $\lambda$ varies from 1.4938 at the origin to 1 at present.
A change of the inertial and gravitational mass is not a new fact, it is well known in Special Relativity, where 
the effective masses change as a function of their velocity.   In the standard model of particle physics, the constant masses of elementary
 particles originate from the interaction of the Higgs field \citep{Higgs14,Englert14} in the 
vacuum with originally massless particles. 
Also in the $\Lambda$CDM, the resulting acceleration of a gravitational system does not let 
its total energy unchanged \citep{Krizek16}. The assumption of scale invariance of  the vacuum  (at large scales) 
 makes the inertial and gravitational masses to slowly slip over the ages, 
however 
 by a rather  limited amount in realistic models (see Fig. \ref{scale}).
 
We  check 
 that the above expression (\ref{consm}) is   fully  consistent with the hypotheses made.
 According to  (\ref{ro2}),  we  have   $\varrho' \lambda^2 = \varrho$, where 
 the prime refers to the value in  GR.
Thus expression (\ref{consm}) becomes,  also accounting for the scale transformation  $ \lambda \, R=R'$, 
$
\varrho_{\mathrm{m}} \, \lambda \, R^3  =  \varrho'_{\mathrm{m}} \, \lambda^3 \, R^3 =  \varrho'_{\mathrm{m}} \, R'^3=const.
$
This is just the usual mass conservation law  in GR. 

Let us go on with the conservation law for relativistic particles and in particular for radiation with density $\varrho_{\gamma}$. 
We have $w=1/3$.  From  (\ref{3w}), we get
$\varrho_{\gamma} \, \lambda^2 \, R^4 =const.$ 
As for the mass conservation, we may check its consistency with GR. This  expression  becomes 
$\varrho'_{\gamma} \, \lambda^4 \, R^4 =const. $ and
thus $\varrho'_{\gamma} \, \, R'^4 =const. $ in  the GR framework. 
Another   case is that of the vacuum  with density $\varrho_{\mathrm{v}}$.  It  obeys to the equation 
of state $ p= -\varrho$ with $c=1$. Thus, we have $w=-1$ and from 
 (\ref{3w}),
$
\varrho_{\mathrm{v}} \lambda^{-2} = const.
$
indicating a decrease of the vacuum energy with time.
With $\varrho'_{\mathrm{v}} \lambda^2 = \varrho_{\mathrm{v}}$, this   gives  $\varrho'_{\mathrm{v}}= const.$ in  the GR framework, 
in agreement with    the standard result. 
The above laws allow us to determine  the time evolution of $\varrho_{\mathrm{m}}$, $\varrho_{\gamma}$ and temperature $T$
 in a given cosmological model, as shown in the next Section.

\section{Numerical  cosmological models}  \label{sect-models}
\begin{figure*}[t!]
\centering
\includegraphics[width=.85\textwidth]{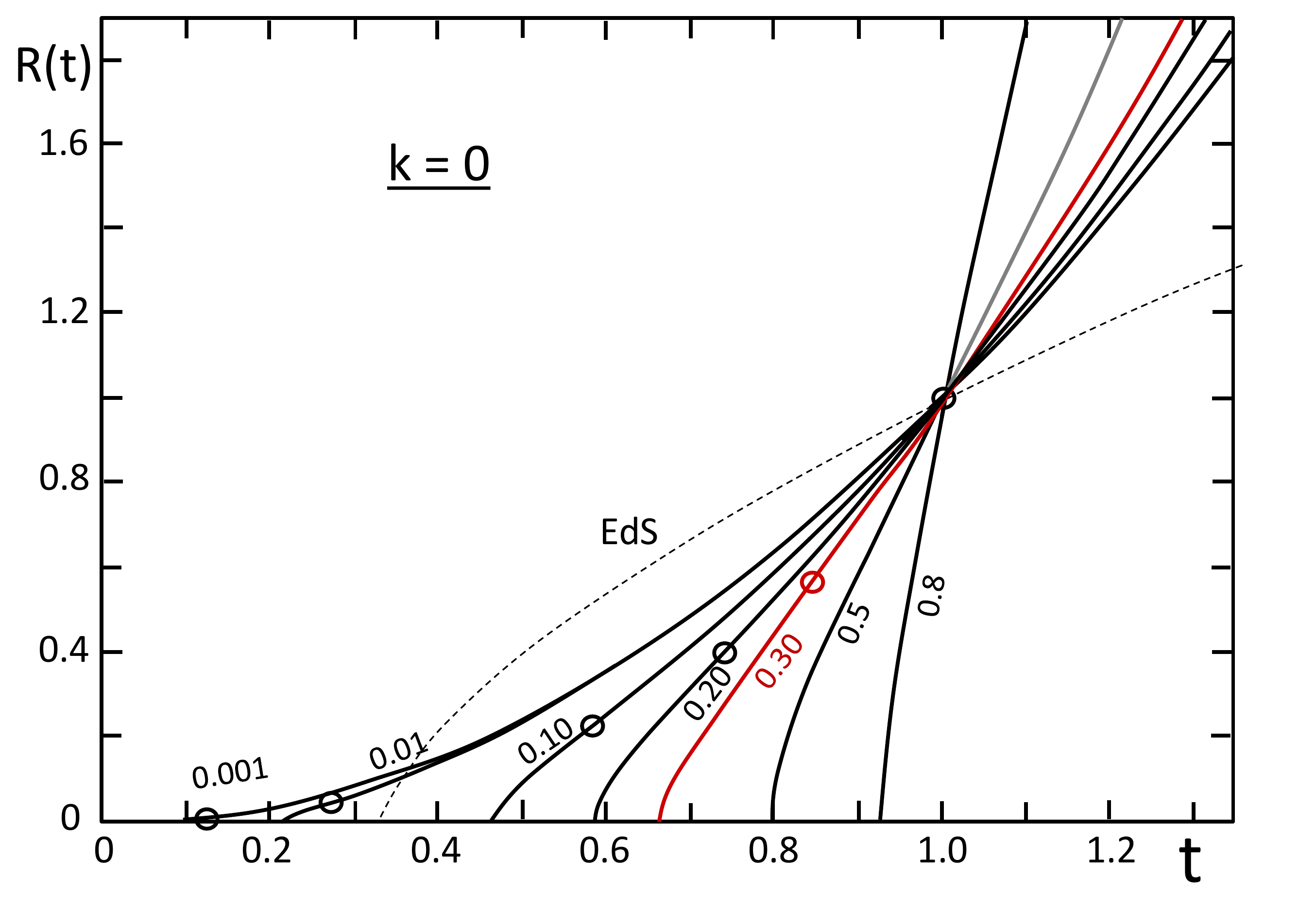}
\caption{Some solutions of R(t) for the models with $k=0$. 
The curves are labeled by the  values of $\Omega_{\mathrm{m}}$, the usual density parameter defined
 in  (\ref{rhocc}). 
 The Einstein-de Sitter model (EdS) is indicated by a dotted line. 
 The small circles on the curves show the transition point between braking
$(q >0) $ and  acceleration $(q<0)$. For  $\Omega_{\mathrm{m}}=0.80$, this point is at $R= 2.52$. 
The   red curve  corresponds $\Omega_{\mathrm{m}}=0.30$ . }
\label{Rtzero}
\end{figure*}
We now construct scale invariant cosmological models and examine their dynamical properties.
The solutions  of  the  equations  are  searched here  for the case of ordinary  matter  
with density  $\varrho_{\mathrm{m}}$
and $w=0$. We may write (\ref{E1}), 
\begin{equation}
\frac{8 \, \pi G \varrho_{\mathrm{m}} R^3 \lambda}{3} = k \, R \lambda +\dot{R}^2 R \lambda+ 2 \, \dot{R} R^2 \dot{\lambda} \, .
\label{E11}
\end{equation} 
\noindent
The first member is a constant. With  $\lambda=t_0/t$ and the present time $t_0=1$, we have
\begin{equation}
\dot{R}^2 R \, t - 2 \, \dot{R}\, R^2 + k \, R \, t - C \, t^2 =0 \, ,  \quad \mathrm{with}
 \quad C=\frac{8 \, \pi G \varrho_{\mathrm{m}} \, R^3 \lambda}{3} \, .
\label{E12}
\end{equation}
\noindent
Time $t$ is expressed in units of  $t_0=1$, at which we also
assume $R_0 = 1$. The origin, the Big-Bang if any one, occurs when $R(t)=0$ at an initial time 
$t_{\mathrm{in}}$ which is not necessarily 0  ($t_{\mathrm{in}}=0$ only for a flat empty  model).
 We  notice that if we have a  solution $R$ vs. $t$,
 then $(x \, R)$ vs. $(x \, t)$ is also a  solution. Thus, the solutions are also scale invariant, 
 as expected from our initial assumptions. 

\begin{table*}[t!]  
\vspace*{0mm} 
 \caption{ {\footnotesize{Cosmological  parameters  of  models with $k=0$ and different $\Omega_{\mathrm{m}} $. 
$H_0(t_0)$  is the present Hubble constant in units of $t_0=1$, $t_{\mathrm{in}}$ is 
the time when $R(t) = 0$, $ \tau= t_0 -t_{\mathrm{in}}$ is the age of the Universe in units of $t_0=1$,
$H_0(\tau)$ is the $H_0$-value in  units of $\tau$, $t$(q) and  $R$(q) are 
the values of $t$ and $R$ at the inflexion point, ``$H_{0\, obs}$`` is the value of $ H_0 $
in km s$^{-1}$ Mpc$^{-1}$.}}
} 
\begin{center} 
\scriptsize
\begin{tabular}{ccccccccccc}
$\Omega_{\mathrm{m}}$  &  $C$  & $H_0(t_0)$ & $t_{\mathrm{in}}$ & $q_0$ &  $\tau$   & $H_0(\tau)$ & 
$t$(q) & $R$(q) & $\Omega_{\lambda}$ &$H_{0\, obs}$  \\
\hline
 &   &   &   \\
0.001  & 0.0040 & 2.0020 & 0.0999 &-0.50 & 0.9001   &1.802 & 0.126 & 0.010 & 0.999 & 127.7\\
0.010  & 0.0408 & 2.0202 & 0.2154 & -0.49 & 0.7846   & 1.585 & 0.271 & 0.047 & 0.990 & 112.3 \\
0.100  & 0.4938 & 2.2222 & 0.4641 &-0.40 & 0.5359    &1.191 & 0.585 & 0.231 & 0.900 & 84.4 \\
0.200  & 1.2500 & 2.5000 & 0.5848 &-0.30 & 0.4152    &1.038 & 0.737& 0.397 & 0.800 & 73.5 \\
0.250  & 1.7778 & 2.6667 & 0.6300 & -0.25 & 0.3700    &0.987& 0.794& 0.481 & 0.750&69.9 \\
0.300  & 2.4490 & 2.8571 & 0.6694 & -0.20 & 0.3306    &0.945 & 0.843 & 0.568 & 0.700 & 66.9\\
0.400  & 4.4444 & 3.3333 & 0.7367 & -0.10 & 0.2633    &0.878 & 0.928 & 0.763 & 0.600 & 62.2\\
0.500  & 8.0000 & 4.0000 & 0.7936 & 0.00 & 0.2064     &0.826 & 1.000 & 1.000 & 0.500 & 58.5 \\
0.800  &  80     &    10    & 0.9282  & 0.30 & 0.0718      &0.718 & 1.170 &  2.520 & 0.200& 50.9 \\
0.990  & 39600  & 200    & 0.9967 &  0.49 & .00335      &0.669  & 1.256   & 21.40  & 0.010& 47.4 \\
\hline
\normalsize
\end{tabular}
\end{center}
\end{table*}
\begin{table*}[htb]  
\tiny
\label{detfin}
\vspace*{0mm}
 \caption{{\footnotesize{Data as a function of the redshift $z$ for  the models with $k=0$  and $\Omega_{\mathrm{m}}=0.30$ (first six
columns)  and $\Omega_{\mathrm{m}}=0.20$ (last four columns).
  Columns 2  and 3 give  $R/R_0$  and
 time  $t/t_0$; column 4 contains the age  in year  with 13.8 Gyr at $t_0$;
column 5 gives the Hubble parameter $H(t_0)$ in the scale $t_0=1$, while column 6 gives  $H(z)$ in km s$^{-1}$ Mpc$^{-1}$ for
$\Omega_{\mathrm{m}}=0.30$. 
The last four columns give $t/t_0$,  the age also with 13.8 Gyr at $t_0$,   $H(t_0)$ and   $H(z)$ for  $\Omega_{\mathrm{m}}=0.20$. }}}
\begin{center} 
 \scriptsize
\begin{tabular}{ccccccccccc}
$z$  & $R/R_0$  & $t/t_0$   & age  &  $H(t_0)$ &    $H(z)$ &  $\quad \quad$ &    $t/t_0$   & age  &  $H(t_0)$ &    $H(z)$   \\
                   &             &     & yr       &             & obs    &                             &          &   yr  &     &    obs        \\
\hline
   &   &   &  \\
  \multicolumn {11} {l}{\bf{ $\quad \quad \quad  \quad$ Models with $\Omega_{\mathrm{m}}=0.30 
 \quad \quad   \quad \quad \quad  \quad \quad \quad  \quad \quad \quad$  Models with  $\Omega_{\mathrm{m}}=0.20$ }} \\
0.00    &   1       &      1       & 13.8E+09     &   2.857  &   67.0   &        &      1       & 13.8E+09     &   2.500   &   73.6  \\
0.10    & .9091   &  .9679    &  12.5E+09    &   3.088   &   72.3  &        &  .9631    &  12.6E+09    &   2.675   &   78.7  \\
0.20    & .8333   &  .9407    &  11.3E+09    &   3.324   &   77.9   &       &  .9316    &  11.5E+09    &   2.852   &   83.9  \\
0.40    & .7143   &  .8974    &    9.5E+09    &   3.810   &   89.3   &       &  .8806    &    9.8E+09    &   3.212   &   94.5  \\  
0.60    & .6250   &  .8644    &   8.1E+09     &    4.321   &  101.2  &      &  .8412    &    8.5E+09    &   3.580   &  105.3  \\
0.80    & .5556   &  .8387    &   7.1E+09     &    4.852   &   113.7 &      &  .8099    &    7.5E+09    &   3.960   &  116.5 \\
1.00    & .5000   &  .8181    &   6.2E+09     &    5.408   &   126.7 &      &  .7845    &    6.5E+09    &    4.352   &  128.0  \\
1.20    & .4545   &  .8013    &   5.5E+09     &     5.987   &  140.2  &     &  .7636    &    5.9E+09    &    4.756   &  139.9 \\
1.50    & .4000   &  .7814    &   4.7E+09       &    6.895   &   161.5 &    &  .7383    &    5.1E+00     &    5.386   &  158.5 \\
1.80    & .3571   &  .7660    &   4.0E+09       &    7.854   &   184.0 &    &  .7184    &    4.4E+09    &    6.045   &  177.9  \\
2.00    & .3333   &  .7575    &   3.7E+09       &    8.522   &   199.6 &    &  .7074    &    4.1E+09    &    6.500   &  191.2 \\
2.34    & .2994   &  .7457    &   3.2E+09       &    9.698  &   227.1  &    &  .6918    &    3.6E+09     &   7.303   &  214.9  \\
3.00    & .2500   &  .7290    &   2.5E+09      &    12.16   &   284.6  &    &  .6694    &    2.8E+09     &   8.963   &  263.7      \\
4.00    & .2000   &  .7131     &  1.8E+09      &    16.24   &   380.5  &     &  .6476    &    2.1E+09     &  11.72   &   344.8   \\
9.00    & .1000   &  .6855    &  6.7E+08     &    42.46   &   994.5    &     &    .6085    &    7.9E+08     &  29.27   &   861.2   \\
\hline
\normalsize
\end{tabular}
\end{center}
\end{table*}

\subsection{The case of a flat space  ($k$ = 0) }  \label{sub:flat}
The case of the Euclidean  space is evidently the most interesting one in view of the confirmed  results of the  space missions
investigating the Cosmic Microwave Background (CMB) radiation with 
Boomerang  \citep{deBern00}, WMAP \citep{Benn03} and the  \citet{Planck14}. Expression (\ref{E12})
becomes
\begin{equation}
\dot{R}^2 R \, t - 2 \, \dot{R}\, R^2  - C \, t^2 =0 \, .
\label{E13}
\end{equation}
\noindent
For $t_0=1$  and  $R_0=1$, with  the Hubble constant $H_0 =\dot{R_0}/R_0$ at the present time, the above relation   leads to
$
H^2_0 -2 \, H_0  = C \, .
$
This gives  $H_0 \, =  \, 1  \pm  \sqrt{1+C}$,
where we take the sign  + since $H_0$ is  positive.
We  now  relate  $C$ to   $\Omega_{\mathrm{m}}$  at time $t_0$. 
We have $\Omega_{\mathrm{m}}= 1- \Omega_{\lambda}$ and thus from (\ref{Omegapr})  
\begin{equation}
H_0 \, = \, \frac{2}{1-\Omega_{\mathrm{m}}} \, .
\end{equation}
\noindent
This  gives $H_0$ (in unit of $t_0$)  directly  from  $\Omega_{\mathrm{m}}$.
We  now  obtain $C$ as a function of $\Omega_{\mathrm{m}}$,
\begin{equation}
C = \frac{4}{(1-\Omega_{\mathrm{m}})^2} -\frac{4}{(1-\Omega_{\mathrm{m}})} =
\frac{4 \, \Omega_{\mathrm{m}}}{(1-\Omega_{\mathrm{m}})^2}  \, ,
\label{copr}
\end{equation}
\noindent
which   allows us to integrate  (\ref{E13})  for a chosen   $\Omega_{\mathrm{m}}$(at present).
As seen above, the scale invariant cosmology with $k=0$, like the $\Lambda$CDM, but unlike the Friedman models, 
  permits a variety of models with different 
 $\Omega_{\mathrm{m}}$. 
This is an interesting property in view 
of the results on the CMB which support a flat Universe \cite{Planck14}.
Eq.(\ref{copr}) shows that for  $\Omega_{\mathrm{m}}$ ranging from $0 \rightarrow 1$, $C$ covers 
the range  from $0$ to infinity.

\begin{figure}[t!]
\begin{center}
\includegraphics[width=10.5cm, height=8cm]{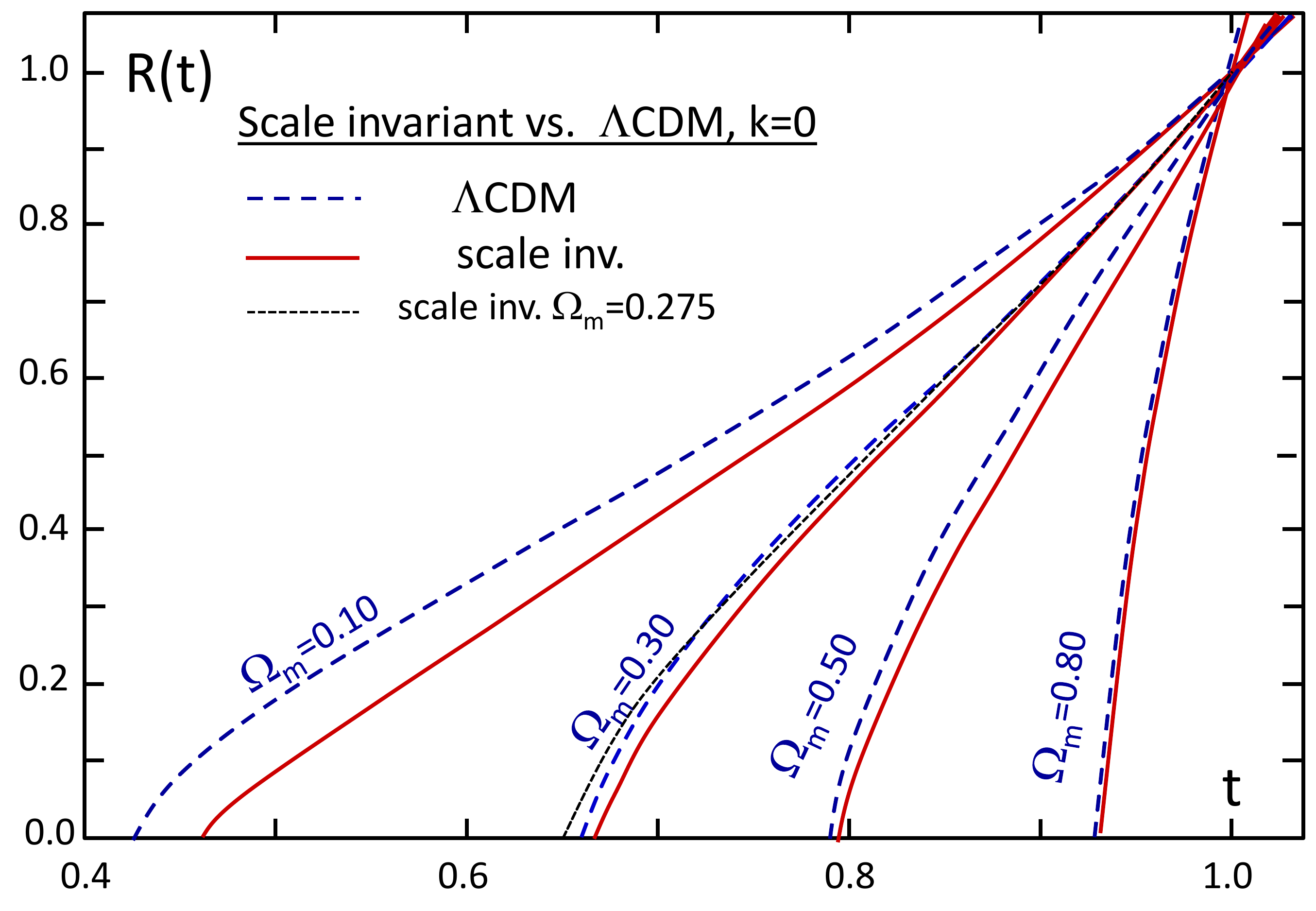}
\caption{{\footnotesize{Comparisons of the $R(t)$ functions of the $\Lambda$CDM and scale invariant models  for given values of
 $\Omega_{\mathrm{m}}$. As an example, the  scale invariant model with $\Omega_{\mathrm{m}}=0.275$ lies very close to
 the $\Lambda$CDM model with  $\Omega_{\mathrm{m}}=0.30$, but the slopes are always slightly different.
}}}
\label{scaLCDM}
\end{center}
\end{figure}
To integrate (\ref{E13}) numerically, we choose a present value for $\Omega_{\mathrm{m}}$, which  determines 
$C$. We proceed to the integration  backwards and forwards in time  starting from  $t_0=1$ and $R_0 =1$.  
Table 1 gives some model data for different $\Omega_{\mathrm{m}}$. The integration provides $R(t)$ and the 
related parameters $H$ and $q$.
Fig. \ref{Rtzero} shows some curves of $R(t)$ for different  values of  $\Omega_{\mathrm{m}}$.
 To get $H_0$ in km s$^{-1}$ Mpc$^{-1}$, we proceed as follows. The inverse of the age of the 
 Universe of 13.8 Gyr  \citep{Frie08} is
  $2.295 \cdot  10^{-18}$ s$^{-1}$, which in the units currently used is equal to 
  70.85 km s$^{-1}$ Mpc$^{-1}$.  This  numerical value   is to be associated  to $H_0(\tau)=1.000$ in column 7
  of Table 1. On the basis of this correspondence,  we multiply all values of $H_0(\tau)$ of
   column 7 by 70.85 km s$^{-1}$ Mpc$^{-1}$
  to get the values of  $H_0$ in current units (last column). The Hubble constant $H_0=  66.9 $  km s$^{-1}$ Mpc$^{-1}$ is  predicted 
  for  $\Omega_{\mathrm{m}} =0.30$ (73.5 km s$^{-1}$ Mpc$^{-1}$  for  $\Omega_{\mathrm{m}} =0.20$).
  The  agreement with recent values of $H_0$ (Sect. \ref{sub:ho})
    indicates that the expansion rate is correctly predicted for the given age of the Universe.

The models of Fig. \ref{Rtzero} show that
 after an initial phase of braking,  there is an accelerated expansion, which goes on all the way. 
No curve $R(t)$ starts with an horizontal tangent, except the empty model,
where the effect of scale invariance are the largest.
All models  with matter start explosively 
with very  high values of $H={\dot{R}} /R$ and a positive value of $q$, indicating  braking.
 The locations  of the inflexion points,  where $q$
changes sign, are indicated  by a small open circle. For $\Omega_{\mathrm{m}}  \rightarrow 1$,  the age of the Universe is smaller. 
At the same time for increasing $\Omega_{\mathrm{m}}$,  $H_0(t_0)$  (in units of $t_0$)  becomes much larger, while
$H_0$ in units of $\tau$ and  in  km s$^{-1}$ Mpc$^{-1}$ becomes  smaller.
 In Table 2, we give some basic data for the reference models  with $k=0$ and $\Omega_{\mathrm{m}}=0.30$ and 0.20
as a function of the redshift $z$. The values of $H(z)$  in 
   km s$^{-1}$ Mpc$^{-1}$ are derived as for Table 1, by the relation
 $ H(z) \, = \, 70.85  \cdot  \tau(z=0) \cdot H(t/t_0).$

Fig. \ref{scaLCDM} shows the comparison of  the scale invariant and
 $\Lambda$CDM models both with $k=0$. The curves 
   of both kinds of models are similar with smaller differences 
for increasing density parameters. The figure shows that the scale invariant model
with $\Omega_{\mathrm{m}}=0.275$ lies  very closely to the  $\Lambda$CDM  model with $\Omega_{\mathrm{m}}=0.30$
with  values of $R(t)$  slightly smaller for $R(t) > 0.30$, and slightly larger for lower values.
These results imply that for  most observational tests the predictions of the scale invariant 
models are  not far  from those of the $\Lambda$CDM models.

\begin{figure}[t!]
\begin{center}
\includegraphics[width=9.5cm, height=7cm]{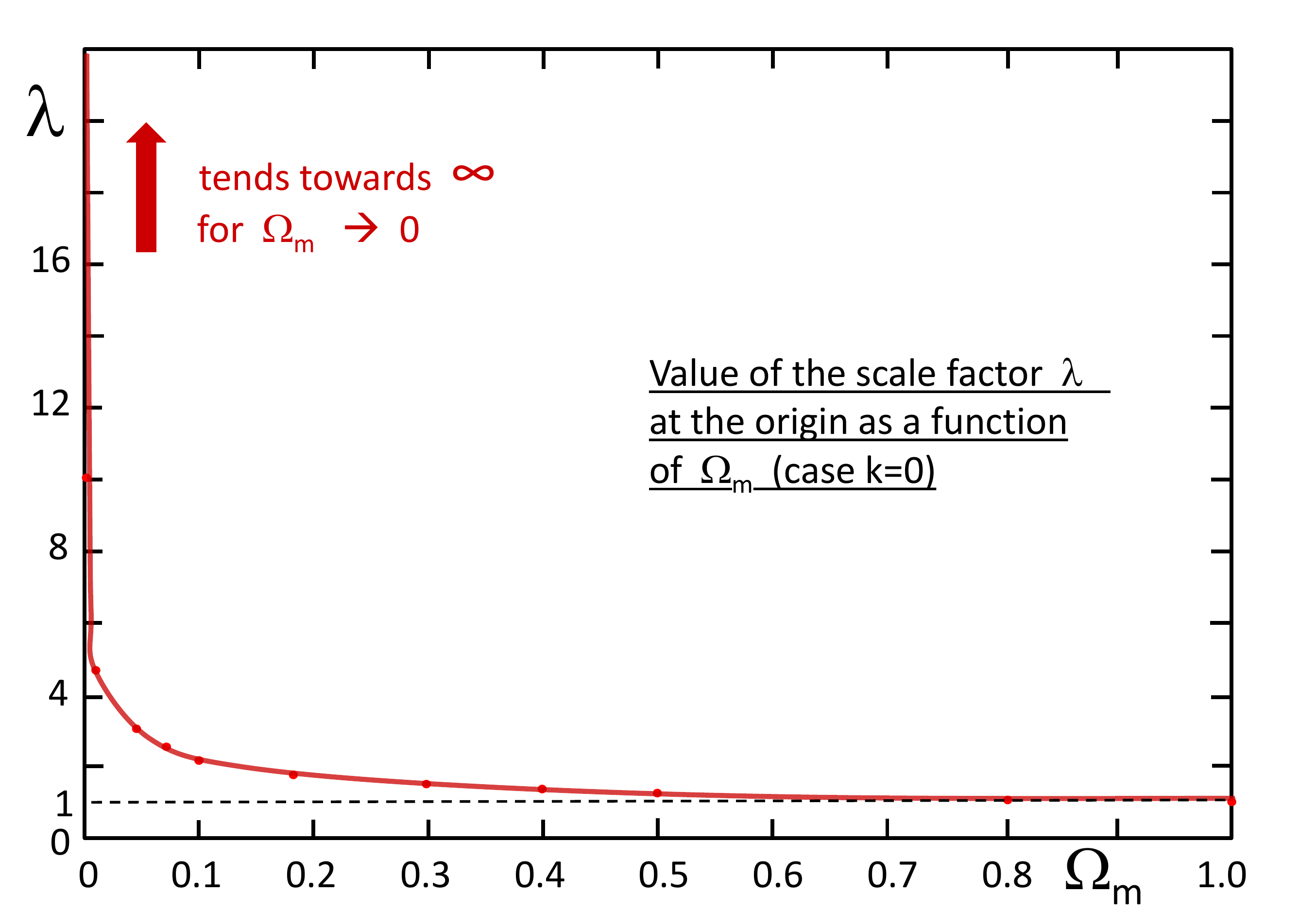}
\caption{{\footnotesize{The scale factor $\lambda$ at the origin $R(t)=0$ for models with $k=0$ and  
different  $\Omega_{\mathrm{m}}$ at present time $t_0$. At $t_0$, $\lambda=1$ for all models. This curve
shows that for increasing densities, the amplitudes of the variations of the scale factor $\lambda$ are very much reduced.
}}}
\label{scale}
\end{center}
\end{figure}
The behavior of $\lambda (t)$ is interesting  (Fig.~\ref{scale}).
 For an empty space, the factor $\lambda$ varies enormously, between $\infty$ at the origin and 1 at present. 
As soon as matter is present, the range of  $\lambda$-values falls dramatically.
For   $\Omega_{\mathrm{m}}= 0.30$, 
  $\lambda$ varies only from 1.4938 to 1.0 between the Big-Bang and  now. Thus,  the 
   presence of less than  1 H-atom per cubic meter
  is sufficient to drastically  reduce the amplitude  of the domain of  $\lambda$-variations.
For $\Omega_{\mathrm{m}} \rightarrow 1$, the scale factor $\lambda$  tends towards  1 all the way. This means 
that the effects of scale invariance  disappear and that the cosmological solutions tend towards those of GR. 
Indeed, 
we have seen that there is no scale invariant solution for $\Omega_{\mathrm{m}}> 1$.
Thus, in the line of the remark by \citet{Feynman63} in the introduction,
  we  see that  the presence of even very tiny amounts of matter in the Universe
 tends to very rapidly kill  scale invariance. The point is that with $\Omega_{\mathrm{m}}\sim 0.30$, the effect appears to be not yet  
 completely killed and may thus deserve the present investigation.

\subsection{The elliptic and hyperbolic scale invariant models}
Although the non-Euclidean models  are not supported by the observations of the CMB radiation \citep{Planck14},
 we briefly present the main 
properties of these models. 
We first have to relate the constant $C$ to    the density  parameters. Expressing $C$ in (\ref{E12}) with
(\ref{rocm1}) and (\ref{Ok}), we get at time $t_0$
\begin{equation}
C= \frac{8 \, \pi \,  G  \varrho_{\mathrm{m}}}{3} = \Omega^*_{\mathrm{m}} \, H^2_0 \, \left(1-\frac{2}{t_0 H_0} \right)  \,  ,
\label{coc}
\end{equation}
\noindent
and with (\ref{relom})
\begin{equation}
 C \,=  \,\Omega_{\mathrm{m}} \, H^2_0 \, .
\label{coco}
\end{equation}
\noindent 
 We  also  have  relation    (\ref{kom}) between the geometrical parameter $k$ 
and  $\Omega^*_{\mathrm{m}}$  at the present time. It allows us to eliminate $[1-2/(t_0 \,H_0)]$  from (\ref{coc}) and  obtain 
\begin{equation}
C \, = \,  \frac{k \, \Omega^*_{\mathrm{m}}}{\Omega^*_{\mathrm{m}}-1}  \, , \quad \mathrm{with} \; \; k= \pm 1  \, .                            
\label{Ck1}
\end{equation}
\begin{figure}[t!]
\centering
\includegraphics[angle=0.0,height=6.5cm,width=9.2cm]{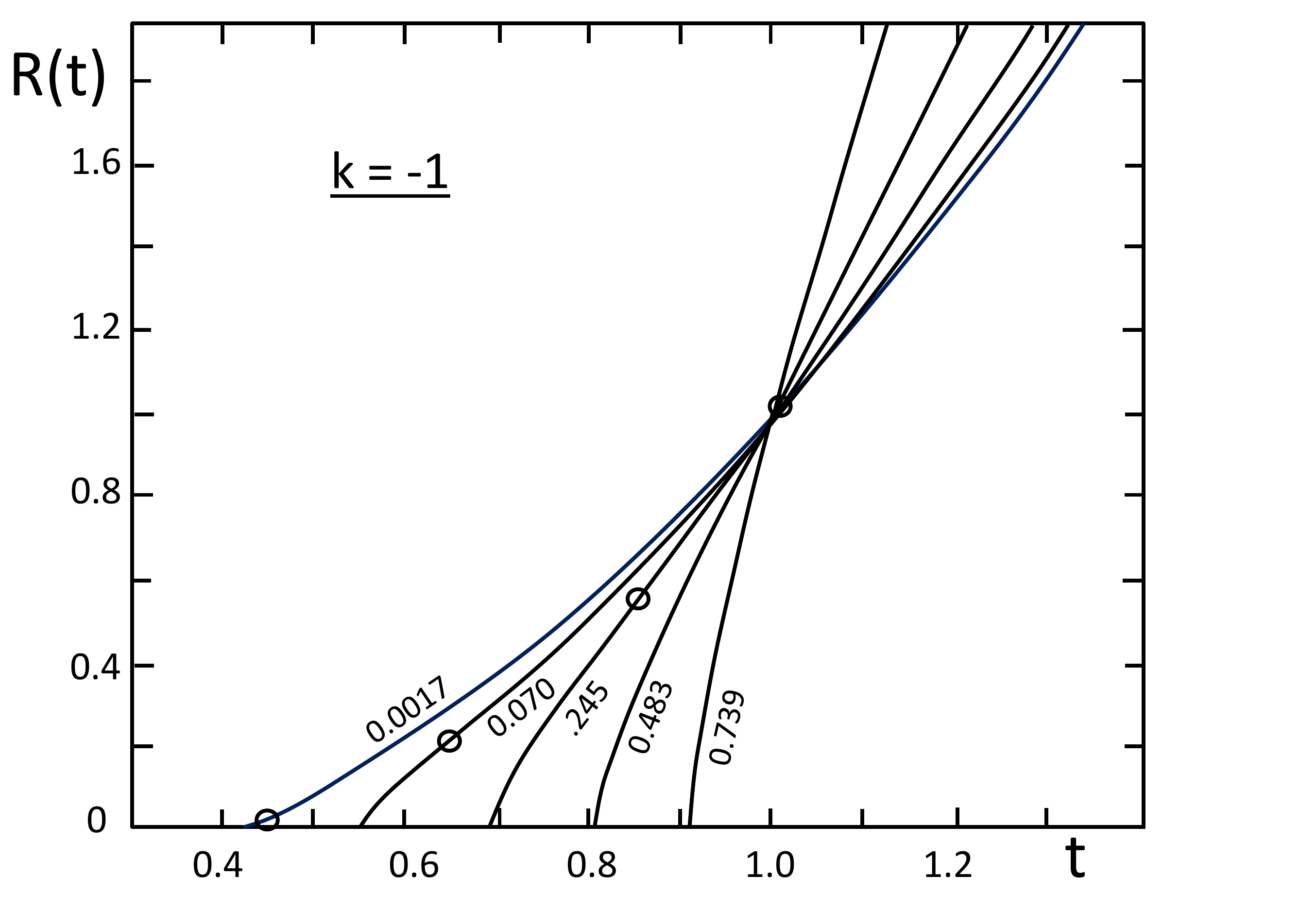}
\includegraphics[angle=0.0,height=6.5cm,width=8.7cm]{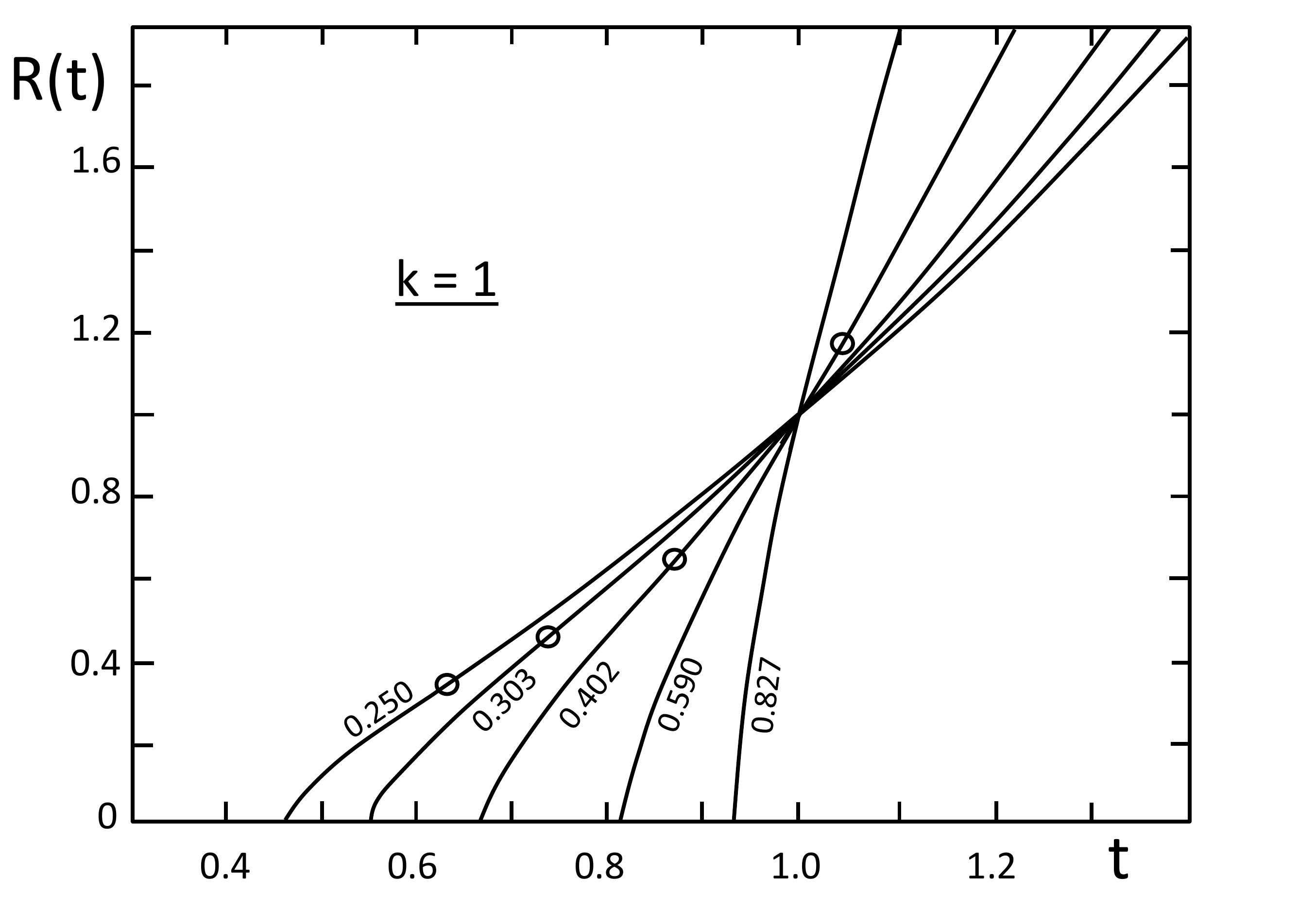}
\caption{Left: Some solutions of $R(t)$ for the models with $k=-1$. The curves are labeled by the  values of 
$\Omega_{\mathrm{m}}$ (at $t_0$), the usual density parameter defined by (\ref{rhocc}). 
The corresponding values of $\Omega^*_{\mathrm{m}}$ used to define $C$ 
are 0.001, 0.315, 0.70, 0.90, 0.98 from left to right. Right: Some solutions of $R(t)$ for $k=+1$. The corresponding
 values of the usual density parameter 
 $\Omega_{\mathrm{m}}$   are indicated. The values of  $\Omega^*_{\mathrm{m}}$ used to define the $C$-values 
 are 1000, 3.0 , 1.5, 1.10, 1.01.
        }
\label{Rt1}
\end{figure} 

\noindent
A model is defined by its $\Omega^*_{\mathrm{m}} $-value at the present time.
For integrating  (\ref{E12}),  we first choose an arbitrary value of  $\Omega^*_{\mathrm{m}}$  for the considered $k$
and then use   (\ref{Ck1})  to obtain the  corresponding $C$--value.  The integration of  (\ref{E12}) from the present  
$t_0 =1$  and $R_0 =1$ is performed forwards and backwards to obtain $R(t)$   and its derivatives. 
The value of $H_0=(\dot{R}/R)_0$  
gives us the  corresponding   $\Omega_{\mathrm{m}}$.
 For non-zero curvature models, $\Omega_{\mathrm{m}} \neq  (1-\Omega_{\lambda})$ at all times and
 $\Omega^*_{\mathrm{m}}$ is not equal to 1 as for $k=0$.   
$\Omega^*_{\mathrm{m}}$ like the other $\Omega$-terms 
 vary with time. Also, all models have  $\Omega_{\mathrm{m}} < 1$ (Sect. \ref{sub:density}).

Figs. \ref{Rt1}  illustrates some solutions for $k= \pm 1$.
 From these figures,  we see that the three families 
of $R(t)$ curves for $k=0$ and $k=\pm 1$ are on the whole not very different from each other.
 The curves for $k=\pm 1$ also show the same succession as for $k=0$  with first a braking  and then an acceleration phase.
 The relative similarity of the three families indicates that the curvature
   term $k$ has a limited effect compared to the  density (expressed by $C$ in the equations)  and to 
the acceleration  resulting from scale invariance. Unlike in the Friedman models, the same density parameters $\Omega_{\mathrm{m}}$
may exist for different curvatures. 

 The models with $k= -1$  (like for $k=0$) have all possible values of $C$ from 0 to infinity, while their  usual density parameter 
$\Omega_{\mathrm{m}}$ remains smaller than 1.0. 
For  $k=+1$, we notice a peculiar  behavior. When $C$ increases from 1 to infinity, 
  $\Omega^*_{\mathrm{m}}$ decreases from infinity to 1.0, while 
 $\Omega_{\mathrm{m}}$ goes from  a limit of 0.25  to 1.0.
 The opposite behavior of the two density parameters  results
from the fact that the term   $[1-2/(t_0 \, H_0)]$  becomes very small. As an example,
 for $\Omega^*_{\mathrm{m}}=1000$,  $H_0= 2.00050$, so that the term $[1 -2/(t_0  \,H_0)] =0.00025$ and
 $\Omega_{\mathrm{m}}=0.25$.   

\section{Some observational properties of scale invariant models}

We examine several observational properties of scale invariant models to see whether there are
 some  incompatibilities resulting from scale invariant models. 
 We   try to concentrate on a few tests 
 where the  observational data   are not derived within some specific cosmological models.


\subsection{Distances vs. redshift in scale invariant cosmology}  \label{sub:dist}

Distances are essential to many cosmological tests. The distance of  a given  object of coordinate $r_1$
  depends on the evolution of the expansion factor $R(t)$.
Important distance definitions  
are those of the proper motion distance $d_{\mathrm{M}}$,  the angular diameter distance
 $d_{\mathrm{A}}$ and the luminosity distance  $d_{\mathrm{L}}$. They are related 
 by
 \begin{equation}
 d_{\mathrm{L}} \,= \,  (1+z) d_{\mathrm{M}} \,=  \, (1+z)^2d_{\mathrm{A}} \, , \quad \mathrm{with}
  \quad d_{\mathrm{M}} \, = \,R_0 \, r_1 \, ,
  \label{distances}
  \end{equation}
  The relations between these three distances are model independent.
 We have 
  \begin{equation}
  R_0 \, r_1 \, = \, \frac{c}{H_0} \int^z_0 \, \frac{dz}{H(z)} \, .
  \label{pmd}
  \end{equation}
  \noindent
  From (\ref{E1}), we may write
  \begin{equation}
  \frac{H^2(z)}{H^2_0} \, = \, \Omega_{\mathrm{m}} (1+z)^3 \, t +
   \Omega_{\mathrm{k}} (1+z)^2 + 2 \frac{H(z)}{H_0} \left(\frac{1}{H_0 \, t}\right) \, ,
   \end{equation}
  \noindent
  where we ignore relativistic particles in the present matter-dominated era. Solving the above relation, we
  get
  \begin{equation}
  H(z) \, = \, \frac{1}{t} \, + \, \sqrt{\frac{1}{t^2}+  H^2_0 \left[  \Omega_{\mathrm{m}} (1+z)^3 \, t +
   \Omega_{\mathrm{k}} (1+z)^2 \right]} \, ,
   \end{equation}
  \noindent
  where the sign  ''+'' has been taken, since  $H(z)$ is positive. This allows one to calculate the various distances (\ref{distances})
  in the scale invariant framework.
   Below we examine carefully one of them, say  the angular diameter distance $d_{\mathrm{A}}=d_{\mathrm{M}}/(1+z)$. 
  \begin{figure}[t!]
\begin{center}
\includegraphics[width=11.5cm,height=8.0cm]{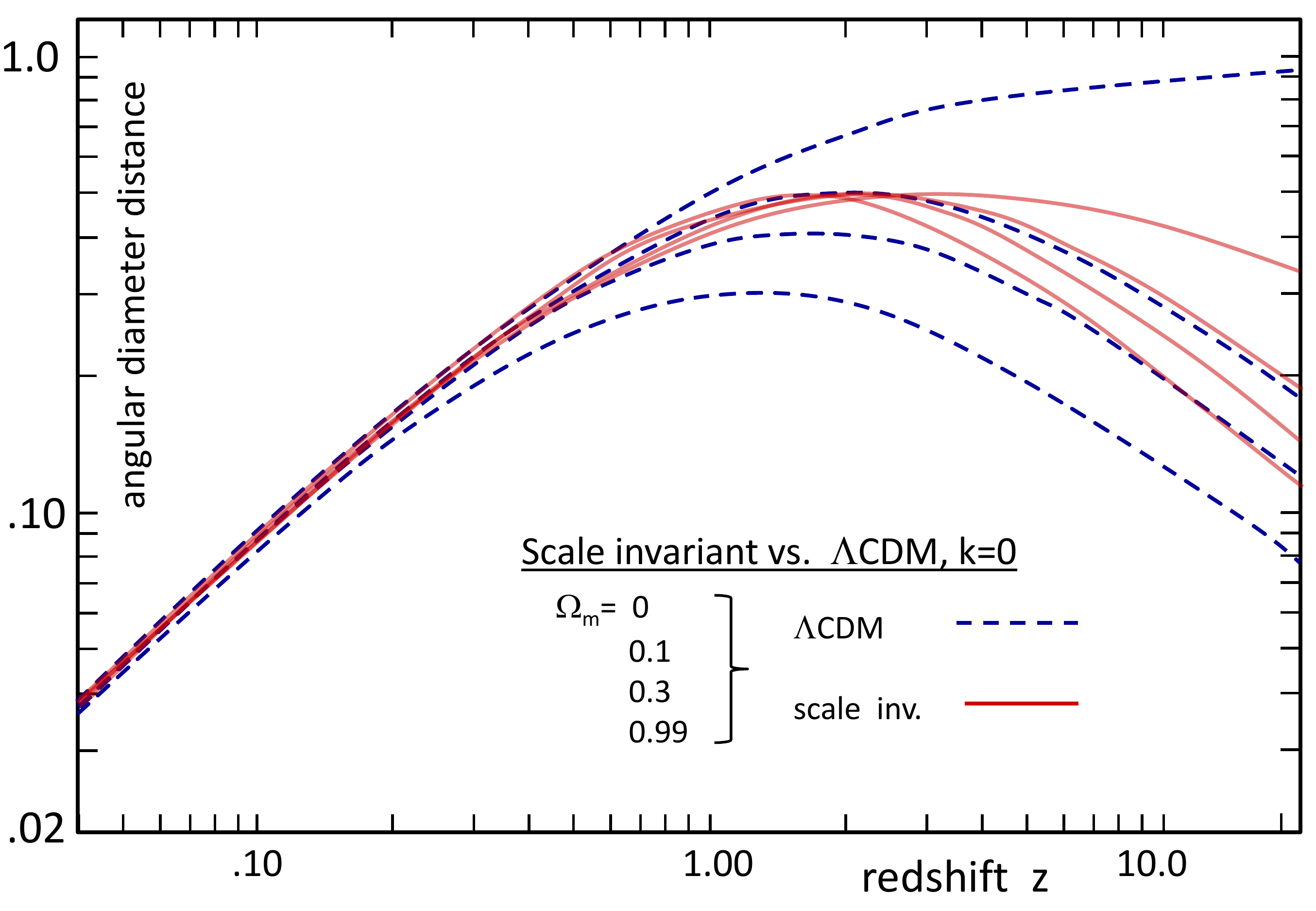}
\caption{{\footnotesize{The angular diameter distance $d_{\mathrm{A}}$ vs. redshift $z$ for flat scale invariant models  (continuous red lines)
compared to flat
$\Lambda$CDM models (broken blue lines).  The  curves are  given for  
$\Omega_{\mathrm{m}} = 0, 0.1, 0.3, 0.99 $,
from the upper to the lower curve  in both cases (at $z  > 3$).}}}
\label{distance}
\end{center}
\end{figure}
Fig. \ref{distance} compares  $d_{\mathrm{A}}$ for the  scale invariant and $\Lambda$CDM models for  $k=0$
and $\Omega_{\mathrm{m}} =  0, 0.1, 0.3, 0.99 $. 
We see that:

-  Up to a redshift $z=2$, the relations between $d_{\mathrm{A}}$ and $z$ for scale invariant models are very close to each other
whatever  $\Omega_{\mathrm{m}}$, with a separation generally smaller than $\sim 0.04$ dex.
At $z=0.6$, for $\Omega_{\mathrm{m}} = 0, 0.1, 0.3, 0.99$, one respectively has
$\log d_{\mathrm{A}}=-0.480, -0.467, -0.450, -0.437$. For $z=1$, the corresponding values are
$-0.383, -0.367, -0.349, -0.342$. 
{\emph {Up to $z \approx 2$, higher  density  models
 lead  to larger $d_{\mathrm{A}}$, while above $z \approx 2$ this is the contrary.}} 
 Then at $z=2$, all curves converge, before slowly diverging for higher $z$.
For $\Lambda$CDM models, higher density models have lower $d_{\mathrm{A}}$
with  an always increasing  separation between the curves.

- At redshift 1, 
the scale invariant models with $k=0$ and $\Omega_{\mathrm{m}}$ between 0 and 0.99  lie between  the 
$\Lambda$CDM models with $\Omega_{\mathrm{m}}$ = 0.1 and 0.3, a trend supported by related tests.

 -  The above properties are evidently also shared by  the proper  motion distances, the luminosity distances, as well as by number
 counts (for $k=0$ at least).
 This means that, if scale invariant cosmology applies,  the cosmological tests
based  on  the magnitude--redshift  diagram, on the angular diameters, as well as on the number counts
up to a redshift of about 2 will require very high precision data to determine 
 $\Omega_{\mathrm{m}}$. A clear discrimination  between the  scale invariant and $\Lambda$CDM models
from tests based  on  distance parameters likely requires redshifts higher than 2.

We note that at high $z$ a fixed angular beam subtends a smaller
scale object for lower density models, giving  thus smaller microwave
background fluctuations \citep{Carr92}. The differences are smaller in the scale invariant case
compared to the $\Lambda$CDM model. Further works will explore such consequences.


\subsection{The magnitude--redshift diagram}
\begin{figure}[t!]
\begin{center}
\includegraphics[width=12.0cm,height=8cm]{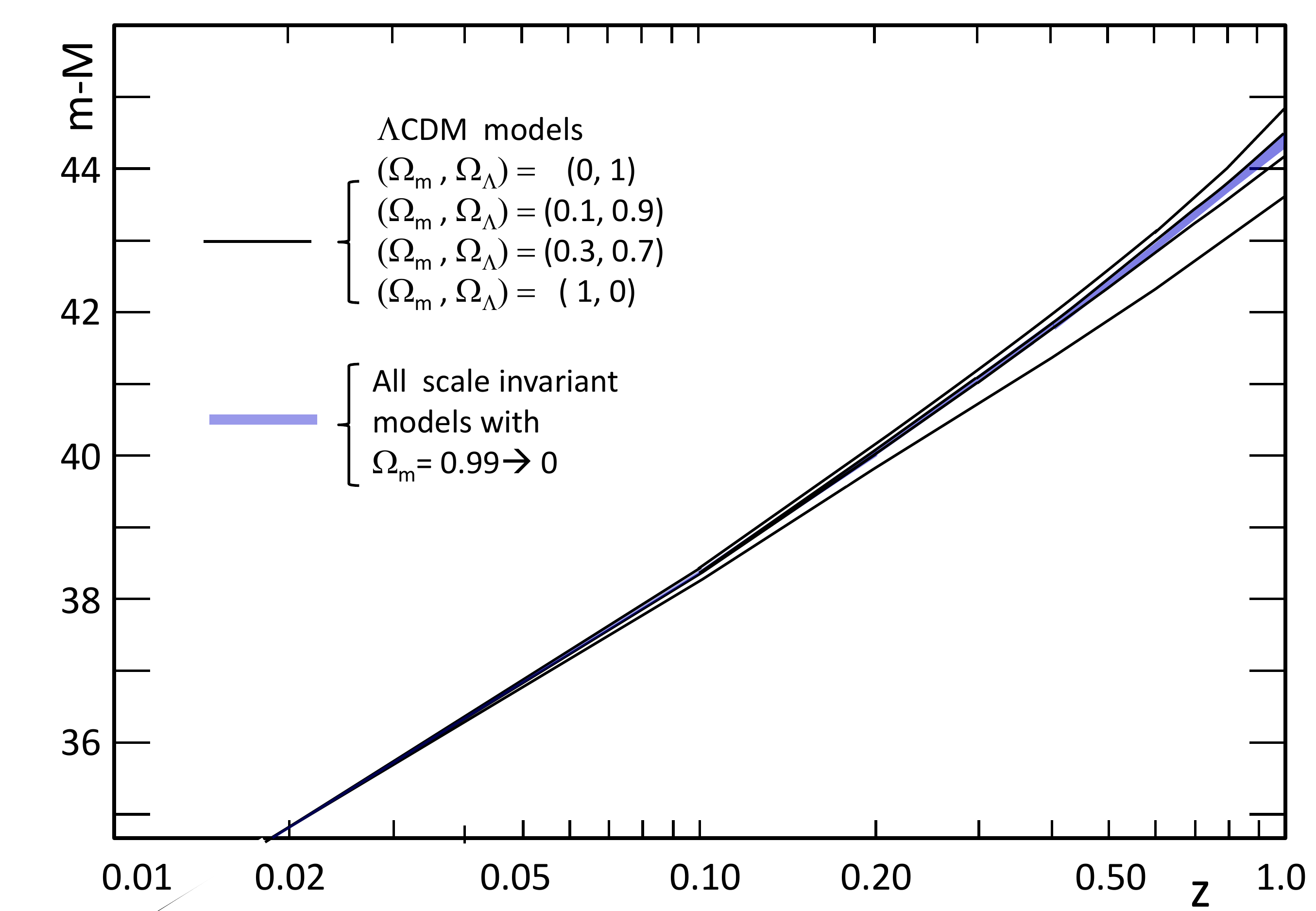}
\caption{{\footnotesize{The magnitude-redshift   diagram for some $\Lambda$CDM  and scale invariant models for
different values of  $\Omega_{\mathrm{m}}$. }}}
\label{mvsz}
\end{center}
\end{figure}
The magnitude-redshift diagram has been a major tool  for  the discovery of the accelerated expansion \citep{Riess98,Perl99}. 
The  flux $f$ received from a distant standard candle, like a supernova of type Ia,  with a  luminosity $L$
goes like  $f = L/(4 \, \pi  \, d^2_{\mathrm{L}})$ and the distance modulus $m-M$ of a source of coordinate $r_1$ is

\begin{equation}
m - M \, = \, const. + 5 \log (R_0 r_1) + 5 \log (1+z) \, , 
\end{equation}

\noindent
where $R_0 r_1$ is obtained from  (\ref{pmd}).
We choose the constant so that  $m-M = 33.22$ at redshift $z=0.01$.
Such an adjustment is  based on the recent  $m-M$ vs. $z$ diagram   from 
 the joined analysis of type Ia supernovae observations obtained in the SDSS-II and SNLS collaborations
given in Fig. 8 by   \citet{Betou14}.
 At this low redshift the various models, whether $\Lambda$CDM or scale invariant, give a similar value of the above constant.
 Relations $m-M$ vs. $z$ are calculated for  a few $\Lambda$CDM and scale invariant models for various
 $\Omega_{\mathrm{m}}$-values. 
 
 Fig. \ref{mvsz} shows some results in the $m-M$ vs. $z$ plot.
 We notice a clear  separation of the curves  for the  $\Lambda$CDM models, which were 
the basic reference   suggesting the accelerated expansion.
   In agreement with the    results for distances, 
   we see that up to  $z=1$ (the present limit in such  plots) the $m-M$ vs. $z$
 curves  show very little dependence on  $\Omega_{\mathrm{m}}$ for scale invariant models.
  Models  with  $\Omega_{\mathrm{m}}$ between 0 and 0.99
 are squeezed within  the thin blue band of Fig. \ref{mvsz}, which  lies between  the curves for $\Omega_{\mathrm{m}}=0.10$
 and 0.30 of the $\Lambda$CDM models.
 In Fig. \ref{mvsz},  the curves of the $\Lambda$CDM models with higher  $\Omega_{\mathrm{m}}$ are lower at all $z$ (higher apparent
 brightness since smaller distance).
 This is also  the case for scale invariant models with $z>2$, but below this redshift the order is inverse with very small separations.
 
%
%
Our main purpose here is not to determine an accurate value of $\Omega_{\mathrm{m}}$ in the scale invariant framework,
but to see whether this theory clashes with current observations. Clearly from the above tests based on the distances, this is not the case.
We note that in the framework of scale invariant models, it may be more difficult to assign
a value of $\Omega_{\mathrm{m}}$ than in the $\Lambda$CDM models due to the smaller effects of the density parameter. 
The $m-z$ diagram for scale invariant  models shows curves  which are not  in contradiction with the current results for
the  $\Lambda$CDM models. These two kinds of models
 may support slightly different  values of $\Omega_{\mathrm{m}}$. 

  \begin{figure}[b!]
\begin{center}
\includegraphics[width=9.0cm, height=7.0cm]{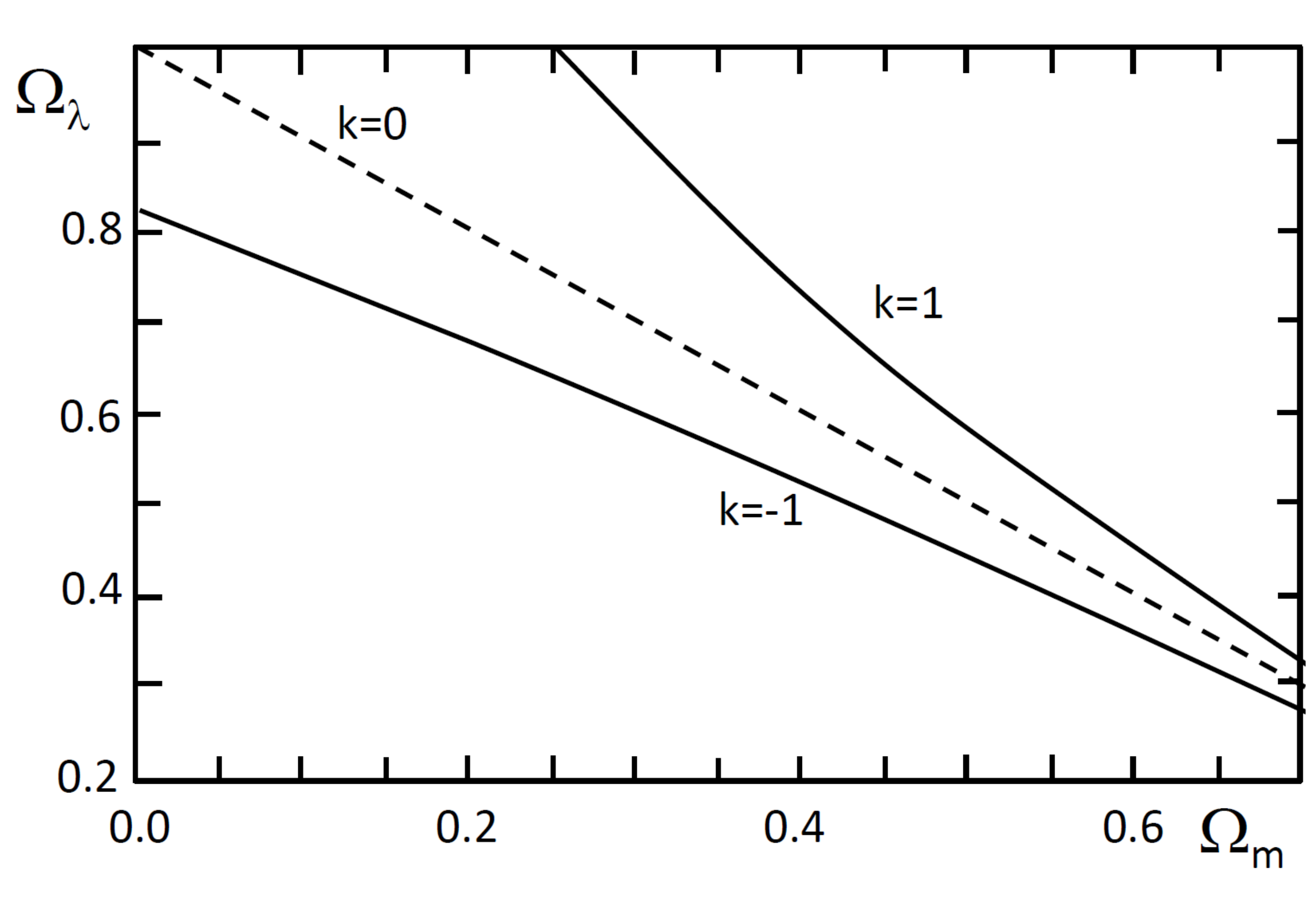}
\caption{{\footnotesize{Plot of the density parameters  $\Omega_{\lambda}$ vs. $\Omega_{\mathrm{m}}$  for scale invariant models 
with $k=0, \; \pm 1$
}}}
\label{omegaml}
\end{center}
\end{figure}
  
  \subsection{ The $\Omega_{\lambda}$ vs. $\Omega_{\mathrm{m}}$ plot}  \label{sub:omlm}
 
The  $\Omega_{\mathrm{m}}$  vs. $\Omega_{\mathrm{m}}$ plot is a major tool in observational cosmology. 
Comparisons between the $\Lambda$CDM models and observational parameters have been performed, 
for example recently, by \citet{Reid10}, \citet{Betou14}, the \citet{Planck14}.  These comparisons 
  clearly favor  values of   $\Omega_{\mathrm{m}}$ around 0.30 and 
$\Omega_{\lambda} \approx 0.70$ in the $\Lambda$CDM models. The point is that the various analyses of
observations, such as the  $m-z$ diagram, the BAO oscillations and the CMB fluctuations, may lead 
to a slightly different value of $\Omega_{\mathrm{m}}$, when interpreted in
the scale invariant models. From the same observations,  different cosmological models lead to different  values of the 
cosmological  parameters.  

 Fig. \ref{omegaml} illustrates the relation between  $\Omega_{\lambda}$ vs. $\Omega_{\mathrm{m}}$ for the various $k
 $-values. The effect of different curvatures is larger for lower density parameters  $\Omega_{\mathrm{m}}$.
  Flat models which are clearly supported by the CMB observations 
 \citep{deBern00,Benn03,Planck14} lie on the  diagonal line $\Omega_{\lambda}=1-\Omega_{\mathrm{m}}$. 
 This is the case for the  preferred $\Lambda$CDM model \citep{Betou14}. 
 The differences  between the scale invariant and $\Lambda$CDM models  being small 
  (cf. Fig. \ref{scaLCDM},  this is not the case for  the EdS
   model in Fig. \ref{Rtzero}),
the resulting point of the flat scale invariant models must lie 
  not so far on the $k=0$ diagonal.
 \begin{figure}[b!]
\begin{center}
\includegraphics[width=11.0cm,height=6.5cm]{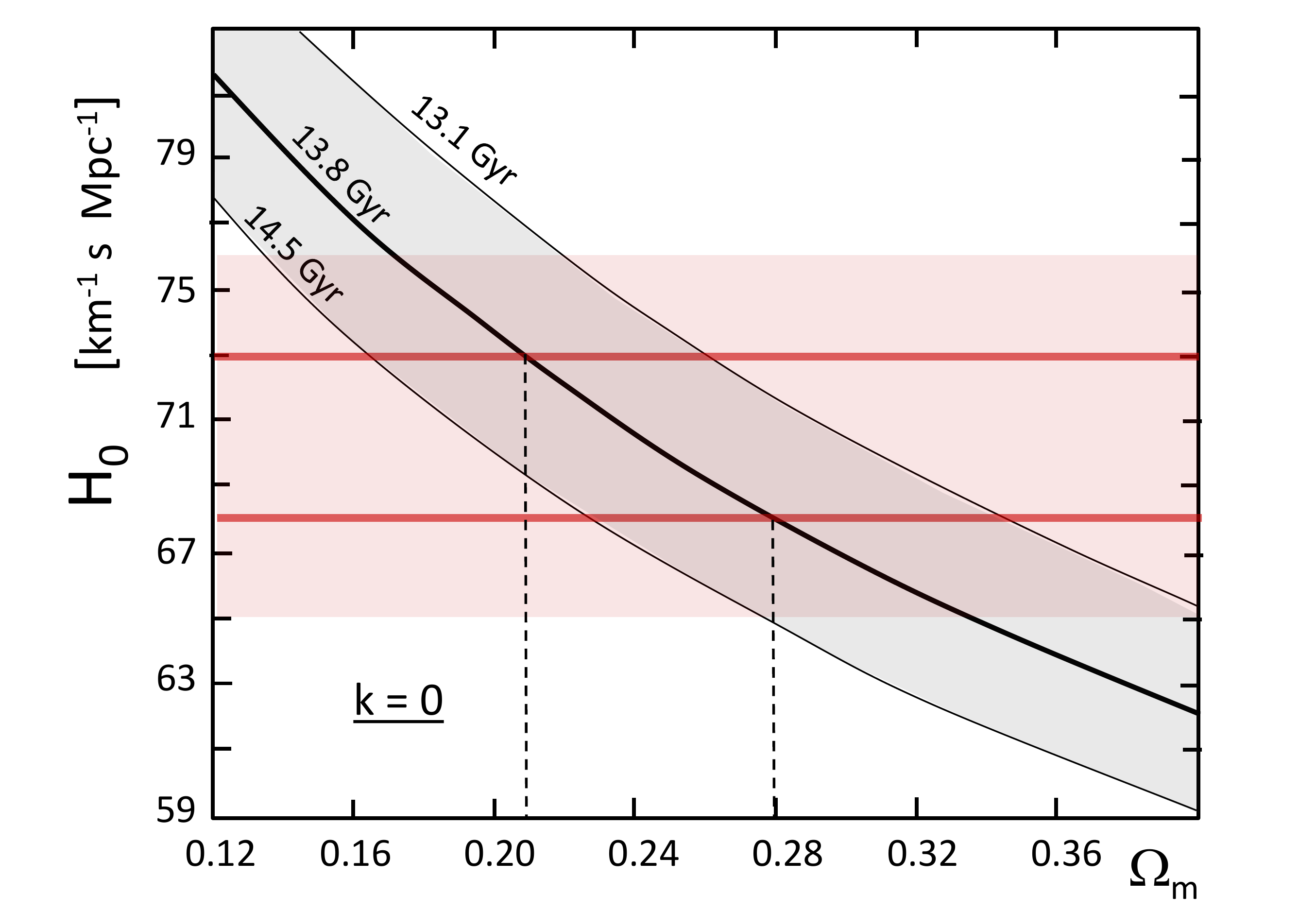}
\caption{{\footnotesize{The relations between  the Hubble constant and $\Omega_{\mathrm{m}}$ for scale invariant models with $k=0$
for  three different values of the age of the Universe (continuous black lines), {\bf{the grey hatched area covers the domain of these values 
corresponding to $\pm 5 \% $ 
differences.  The  low and high values of $H_0$ discussed in the text are 
indicated with an uncertainty band  (pink area) of $\pm$  3  km s$^{-1}$ Mpc$^{-1}$.}}}  }}
\label{HoOmega}
\end{center}
\end{figure}

 \subsection{The relation between $H_0$, $\Omega_{\mathrm{m}}$ and the age of the Universe }  \label{sub:ho}

 An important  test  independent on cosmological models is provided by  the value of the Hubble constant  $H_0$.
 For a given age of the Universe, the value of $H_0$ in km s$^{-1}$ Mpc$^{-1}$
 depends on $\Omega_{\mathrm{m}}$, as shown by Table 1. Thus, for an age of the Universe of 13.8 Gyr (with some
 uncertainty limits),
 we get a constraint on   $\Omega_{\mathrm{m}}$ from the observed value of  $H_0$ (Fig. \ref{HoOmega}).

There has always been some scatter in the results for  $H_0$,  
a scatter which has remarkably decreased with time.
%
%
{\bf{There is still today a bi-polarization in the results for $H_0$, with on one side   the values based on the distance ladder which support a value of 
$H_0$ around 73 km s$^{-1}$ Mpc$^{-1}$.  In particular, \citet{Frie08} gave a value $H_0= 72 \pm 5 $ (in the same units),
\citet{Freedman10}  
 obtained a value   $H_0 = 73  \pm 2$ (random)  $ \pm 4$   (systematic).  Recently,  \citet{Riess16} proceeded to a new 
 determination based on HST data   
 and  obtained  $H_0 = 73.00 \pm 1.75$. 

The above values  appear as outliers when compared to the accumulation of lower values around $H_0 =  68$ km s$^{-1}$ Mpc$^{-1}$.
  Noticeably, the median of a collection of 553 determinations of the Hubble constant converges towards the low value, giving
  $H_0=68 \pm 5$ km s$^{-1}$ Mpc$^{-1}$   \citep{ChenRatra11}. 
  \citet{Sievers13} from high-$\ell$ data for  the CMB power spectrum from
 the Atacama Cosmology Telescope combined with data from WMAP-7yr support  a value $H_0=70
 \pm 2.4$ within the $\Lambda$CDM model. This is consistent with the WMAP-9 yr value of 
 $H_0=70.0 \pm 2.2$  ($69.33 \pm 0.88$ when combined with BAO data) derived by
  \citet{Hinshaw13}. Also, the \citet{Planck14} gives $H_0 = 67.3 \pm 1.2$ within the six-parameter
  $\Lambda$CDM cosmology.
     The combination of BAO and SN Ia data into an inverse distance ladder
  leads  to a value $H_0 = 67.3 \pm 1.1$  \citep{Aubourg15}. This value is independent on the
  $\Lambda$CDM model and rests only on the value of the standard ruler used in BAO analyses.
  This remark also applies to the results from BAO data combined with SN Ia data from JLA
  by \citet{L'Huillier15}, who obtain  $H_0=68.49 \pm 1.53$. On the basis  of 28 measurements
  of the Hubble parameter $H(z)$ analyzed by  \citet{Farooq13}, \citet{Chen16} also support a low value of $H_0= 68.3 \; (+2.7,
  -2.6)$. They show that low values of $H_0$ are also obtained with different assumptions for the 
   dark energy models. The problem of the two separated groups of  favored values is also emphasized by \citep{Bernal16}, who investigate 
   some possible  ways in the early-time physics to reconcile   the high values from the cosmic distance ladder 
    with the low ones more generally found.

Fig. \ref{HoOmega} shows the relation  between
the predicted $H_0$ and  $\Omega_{\mathrm{m}}$ for a reference age 
of 13.8 Gyr (thick black line) in the scale invariant models, and for ages differing by $\pm 5\%$, 
(other  uncertainty limits may easily be drawn). Both the low and high values of $H_0$ are considered here.
A value of $\Omega_{\mathrm{m}}=0.21$  corresponds to the observed $H_0=73$
km s$^{-1}$ Mpc$^{-1}$, while  the 5\% limits lead to a range of $\Omega_{\mathrm{m}}= 0.165$  to 0.26.
A value of $H_0=68 $ km s$^{-1}$ Mpc$^{-1}$ leads to $\Omega_{\mathrm{m}}=0.28$,  the $\pm 5\%$ limits
corresponding to about  $\Omega_{\mathrm{m}}= 0.23$  and 0.345.
Thus, the low $H_0$ value leads to a density parameter $\Omega_{\mathrm{m}}$ in rather agreement with the current values,
while the high $H_0$ value  leads to a lower  $\Omega_{\mathrm{m}}$   than  usually obtained.

 On the whole, the conclusion
we may  draw here is that,  even  if the scale invariant theory   leads to slightly 
different   estimates of $H_0$  and  $\Omega_{\mathrm{m}}$ from those of the  $\Lambda$CDM models,
 it leads to no significant contradiction with current estimates.}}


\subsection{The history of the expansion }   \label{sub:dyna}

The key prediction of the basic equations of cosmological models concerns the expansion function  $R(t)$ of the Universe. 
{\bf{It has an effect on  the distances and  related tests as shown above, as well as on the present and past expansion rates.

\subsubsection{The past rates  $H(z)$ vs. redshifts $z$} \label{subsub:hz}  }}

 The expansion rates $H(z)$ vs. $z$ represent a 
 direct and constraining  test on $R(t)$ over the ages.
 Moreover, this test involves much higher redshifts than the classical $m-z$ diagram.
In order to perform valuable tests, it is  necessary that the observational 
data are  independent on the cosmological models.

%
%

{\bf{Some results on $H(z)$ are obtained by  the method of the cosmic chronometer, which  contains no assumption 
depending on a particular cosmological  model 
 \citep{Jim02,Simon05,Stern10,Moresco15,Moresco16}.   
  This  method  is based on the simple relation 
\begin{equation}
H(z) =- \frac{1}{1+z} \, \frac{dz}{dt} \, ,
\label{hz}
\end{equation}
\noindent
obtained from $R_0/R=1+z$ and the definition of $H=\dot{R}/R$. The critical
ratio $dz/dt$ is estimated from  a sample of passive galaxies 
(with ideally no active star formation) of different redshifts and age estimates.  
 We note, however, that the method of cosmic chronometers, although independent on the cosmological models, 
 depends on the models of spectral evolution of galaxies, mainly based on the theory of stellar evolution.
The detections of BAO in  surveys of high redshift quasars also provide powerful
 constraints. This is  in particular the case for a study  based on 48 640 quasars  in the redshift
range of $z=2.1$ to 3.5  by \citet{Busca13} from the BOSS survey in the SDSS-III results. This study has been 
extended to encompass 137 562 quasars in the same redshift range from the BOSS data release 
DR11 \citep{Delubac15}, (since the second study is an enlargement of the first one, below we only consider the second set of data).
The comparisons between observations and models we perform are using the useful collection  of 28 Hubble parameters  
$H(z)$  by \citet{Farooq13} based on
 \citet{Simon05}, \citet{Stern10}, \citet{Moresco12}, \citet{Busca13}, \citet{Zhang14}, \citet{Blake12} and \citet{Chuang13}.
These data are  completed here by three other recent high precision results by
 \citet{Anderson14}, \citet{Delubac15} and \citet{Moresco16},  they are shown in red color in Fig. \ref{history}.
 
It is advantageous to perform the comparisons   with the ratio $H(z)/(z+1)$ as a function of $z$ rather than with $H(z)$ vs. $z$,
(I am indebted to the referee for this remark). 
 The deceleration parameter $q$  (Sect. \ref{sub:geometry}) can be written,
\begin{equation}
q=-\frac{\ddot{R} R}{\dot{R}^2}= -\frac{dH}{dz} \frac{dz}{dt} \frac{1}{H^2} -1 = \frac{dH}{dz} \frac{1+z}{H}- 1 \, ,
\label{qz}
\end{equation}
\noindent
where  (\ref{hz}) has been used.
If $dH/dz$ decreases with increasing $z$,  the system accelerates and consistently with (\ref{qz}) the $q$--parameter is negative.
Then,  the derivative   
\begin{equation}
\frac{d}{dz}\left(\frac{H(z)}{1+z}\right) =\frac{1}{1+z} \left(\frac{dH}{dz} - \frac{H(z)}{1+z}\right) \,
\end{equation}
\noindent
is negative and the considered ratio is decreasing. The minimum of  $H(z)/(z+1)$ is given by 
$\frac{dH}{dz} = \frac{H(z)}{1+z}$
which also implies that $q=0$. Thus, the  minimum of the considered function occurs at the transition redshift  from braking
to acceleration. Then, for higher $z$, the braking means positive values of   $q$ and of $dH/dz$, implying that   $H(z)/(z+1)$
is an increasing function. These properties makes $H(z)/(z+1)$ a very useful function. 

 Fig. \ref{history} compares the plot of the ratios  $H(z)/(z+1)$ ratios vs. $z$  from \citep{Farooq13}  and complements 
 for  various models: the Einstein-de Sitter (EdS) model,
 the $\Lambda$CDM and the  scale invariant models
 for $k=0$ and $\Omega_{\mathrm{m}}=0.30$.  We emphasize that the results also depend on the assumed age of the Universe,
 since it fixes the numerical value of $H_0$ in km s$^{-1}$ Mpc$^{-1}$ for the model with $H_0(\tau) =1.0$ (see Sect. \ref{sub:flat}).
As discussed above, we generally adopt an age of 13.8 Gyr. In Fig. \ref{history}, we also show the effects for an age larger by $3 \%$
than the standard value. This corresponds to age of about 14.2 Gyr, well within the uncertainty domain of 13.8 $\pm 0.6$ Gyr 
of the consensus model by \citet{Frie08}. From Fig. \ref {history}, we note the following model properties:
\begin{figure}[t!]
\begin{center}
\includegraphics[width=13.5cm,height=10.5cm]{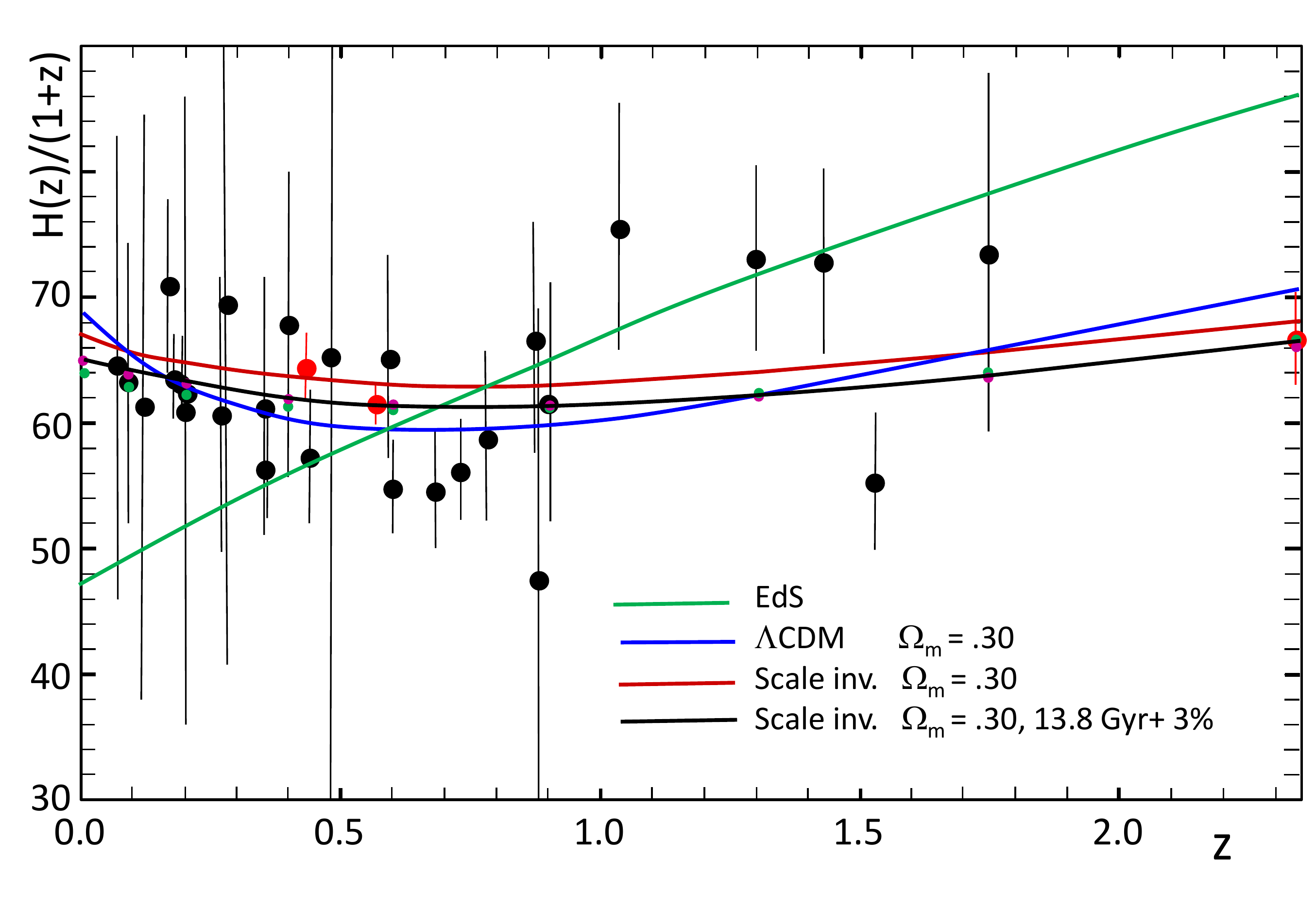}
\caption{{\footnotesize{{\bf{The plot of the  $H(z)/(z+1)$ ratios vs. redshifts $z$ with $H(z)$ in km s$^{-1}$ Mpc$^{-1}$
reproduced from \citet{Farooq13}.
The observations, represented by black dots, show the  data collected by \citet{Farooq13}, 
with the indications of the  $1 \sigma$ uncertainty (see text). Three  other recent 
  values  are added as red points. The  red point at z=2.34 is from the BAO data of the
 BOSS DR11 quasars by \citet{Delubac15},  at z=0.57  it is from 
 \citet{Anderson14} and at z=0.43  from \citet{Moresco16}.
 The green line corresponds to the Einsten-de Sitter model (EdS). 
 The  blue line  shows  the $\Lambda$CDM for 
 $k=0$ and  $\Omega_{\mathrm{m}}=0.30$ and an age of 13.8 Gyr.
 The  red line  corresponds to
 the best fit scale invariant model for same parameters.
  The  black line shows  the same model as the red line, but for an age $3\%$ larger, i.e. of 14.2 Gyr. }} }}}
\label{history}
\end{center}
\end{figure}
 \begin{enumerate}
 
  \item For the EdS model, the continuous  braking makes $H(z)/(z+1)$ to always  increase and to be in clashing
   contradiction with the observations,
 especially more than the value of $H_0$ at $z=0$ is very low. Indeed, for this model $H_0= (2/3) (1/t_0)$.
  For $t_0= 13.8$ Gyr, we have  $1/t_0= 70.85$ km s$^{-1}$ Mpc$^{-1}$ in the considered units
 and taking the $2/3$ leads to 47.23 km s$^{-1}$ Mpc$^{-1}$.  This is the well known age problem of the EdS model.
 
 \item The values of $H_0$ for the $\Lambda$CDM and scale invariant models are also fixed on the basis of an age of 13.8 Gyr
 and of the expansion rate of the models. The facts that both models give consistent values of $H_0$ is a valuable point
 for these two  models. The scale invariant model
 gives $H_0= 66.9$ km s$^{-1}$ Mpc$^{-1}$   (cf. Table 1) and the $\Lambda$CDM model 68.3 km s$^{-1}$ Mpc$^{-1}$, both for $k=0$
 and $\Omega_{\mathrm{m}}=0.30$.

 \item The depth of the minimum of the curve $H(z)/(z+1)$ is more pronounced for the $\Lambda$CDM model, 
 with a difference  up to about  4  km s$^ {-1}$ Mpc$^{-1}$
    compared to the scale invariant model, although the redshifts of the minimum 
    are  close to each other for  $\Omega_{\mathrm{m}}=0.30$ (see Sect. \ref{transition}).
    
    \item  For $z \geq 1.7$, the  ratio  $H(z)/(z+1)$ of the 
    $\Lambda$CDM model becomes significantly higher than for the scale invariant  case. This may provide valuable tests in future.

\end{enumerate}    }}

 In this context, we note that several authors have found some tension between the $\Lambda$CDM models and the $H(z)$
 observations. First, \citet{Delubac15} point out  a 2.5 $\sigma$ difference  at $z=2.34$
 with the predictions of a flat $\Lambda$CDM model. A strong tension is further emphasized by \citet{Sahni14} and
 by \citet{Ding15}, who suggest that ''allowing dark energy to evolve seems to be the most 
 plausible approach to this problem''. 
 \citet{Sola15} find a significantly better 
 agreement with a time-evolving $\Lambda$ depending on $H^2$ and $dH/dt$.
The fundamental constant nature of $\Lambda$ is further questioned by
  \citet{Sola16}. {\bf{We will see below that the good agreement of both the $\Lambda$CDM and scale invariant models
  with  observations lets this question open.}}

\subsubsection{ Comparisons of models and observations. The $\chi^2$ test.}

{{\bf{Here, we examine whether  scale invariant models agree with  observations
and  at the same time we may also see whether  the claims about disagreements between 
observations and the $\Lambda$CDM model are further supported.

The 30 observations collected above are shown  in Fig. \ref{history} with the indications
of the 1 $\sigma$ scatter and are compared to a few model curves. At the eye inspection,
it is difficult to make a selection, apart from the EdS model. Thus, we proceed to some  $\chi^2$ tests.
Firstly, we consider the $\Lambda$CDM model with $k=0$ and $\Omega_{\mathrm{m}}=0.30$, which is of interest for 
comparisons with the scale invariant models below and is also very close to the best fit model
by \citet{Farooq13}  with $\Omega_{\mathrm{m}}=0.29$ or 0.32 (depending on $H_0$). For the sample considered, we find 
$\chi^2= 23.17$.  If we remove the values at $z=0.43$ and 0.57 and replace the new data at $z=2.34$ by the former one 
at $z=2.30$, our sample  is the same as that by \citet{Farooq13}.   It thus leads  to  $\chi^2= 18.88$ which is quite consistent, owing to the 
sightly different  $\Omega_{\mathrm{m}}$, with the minimum values of 18.24 and 19.30 found by these authors. 
These low $\chi^2$ results confirm 
the good agreement of the $\Lambda$CDM model and observations.
In Fig. {\ref{history}, we also compare the flat scale invariant model with $\Omega_{\mathrm{m}}=0.30$
 for an age of 13.8 Gyr to the observations. This model is, however, not the model realizing the best fit with observations.
There,  $\chi^2= 26.87$, which is higher that the value of 23.17,
 obtained from the $\Lambda$CDM model for the same sample.  

We get the best fit  for models with an age  higher by 3\%  (14.2 Gyr) than the above value.
Over the range  $\Omega_{\mathrm{m}}=0.30$  to 0.34, the differences are  insignificant. 
For $\Omega_{\mathrm{m}}=0.30$, we get   $\chi^2= 20.49$,  a satisfactory value.
 We notice that   the $\Lambda$CDM  with  $\Omega_{\mathrm{m}}=0.30$ appears to better reproduce  the low values of $H(z)/(z+1)$ 
    between about $z=0.4$  and 0.9, while the scale invariant with the same $\Omega_{\mathrm{m}}$ 
    does better for observations with $z \geq 1.0$.      This shows the great interest of future observations.  
However,  owing to the scatter of the data, it likely not very meaningful to enter into closer comparisons 
in order to disprove or support one particular model.
  We just conclude that the scale invariant models like the $\Lambda$CDM models, give both consistent results with the observations. 
   Thus,  scale invariant models  may
  provide a possible alternative to the $\Lambda$CDM models and further tests are needed.  }}}

  \subsection{The transition redshift from braking to acceleration} \label{transition}

\begin{figure}[t!]
\begin{center}
\includegraphics[width=14.0cm, height=9.0cm]{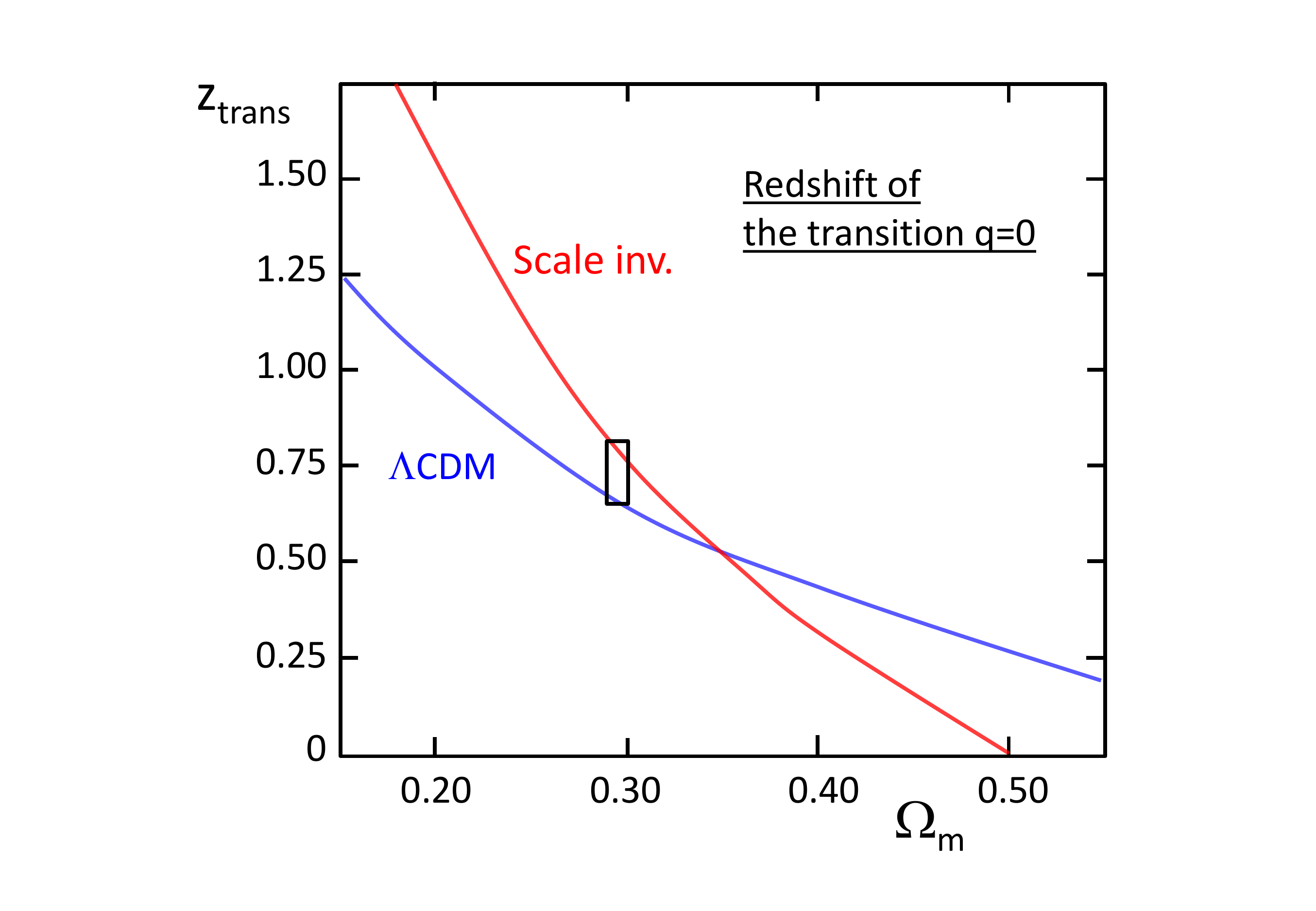}
\caption{{\bf{Relation between the redshift of the transition from  braking to  acceleration 
 vs. the matter density $\Omega_{\mathrm{m}}$ for the flat $\Lambda$CDM and scale invariant models.
The observational values mentioned  in the text  are
within the small black rectangle.}} }
\label{ztrans}
\end{center}
\end{figure}

{\bf{
We have seen in Sect. \ref{sub:inflex} the conditions for the occurrence of the
transition from braking to acceleration which produces an inflexion point
 in the expansion $R(t)$. For the scale invariant model with $k=0$, $q=0 $ occurs
when  $\Omega_{\lambda}= \Omega_{\mathrm{m}} = \frac{1}{2}$.
 For $ \Omega_{\mathrm{m}} =0.30$, the transition occurs at $R/R_0= 0.568$ (cf. Table 1) corresponding to
a transition redshift $z_{\mathrm{trans}}=0.76$. In the $\Lambda$CDM model, the transition lies at  \citep{Suther15},
\begin{equation}
1+z_{\mathrm{trans}}=\left( \frac{2\, \Omega_{\Lambda}}{\Omega_{\mathrm{m}} }\right)^{1/3} \, ,
\end{equation}
\noindent
so that for the same $\Omega_{\mathrm{m}} $, one has $z_{\mathrm{trans}}=0.67$, i.e. it occurs  slightly later in the expansion.
In both kinds of models,  the transition  from braking to acceleration is not a sharp 
one (cf.  Fig. \ref{scaLCDM}), the two phases being separated 
by a non negligible transition phase where $R(t)$ is almost linear.

Several authors have tried to estimate the value of $z_{\mathrm{trans}}$.   The data by \citet{Farooq13}, that we have extensively used above,
enabled these authors to estimate a transition redshift $z_{\mathrm{trans}}=0.74 \pm 0.05$. This value is very well supported by 
other recent ones. \citet{Busca13} gave a value of  0.82 $\pm 0.08$. \citet{Blake12} found $z_{\mathrm{trans}} \approx 0.7$,
the same value is also supported by \citet{Suther15} and by \citet{Rani15}. The best fit by   \citet{Viten15}  supports a transition 
redshift  $z_{\mathrm{trans}} \approx 0.65$. \citet{Moresco16} find a value 
 $z_{\mathrm{trans}}=0.4 \pm 0.1$ for one  model of spectral evolution and
  $z_{\mathrm{trans}}=0.75 \pm 0.15$ for another model.

 Fig. \ref{ztrans} shows as a function of    $\Omega_{\mathrm{m}} $ 
the redshifts $z_{\mathrm{trans}}$ at which the transitions are  located for both the $\Lambda$CDM and the scale invariant models.
   The value of $z_{\mathrm{trans}}$  varies faster with matter density for the scale invariant 
 than for  the $\Lambda$CDM case. However,  the two curves are crossing at about 
 $\Omega_{\mathrm{m}} \approx 0.35$ so that they are still rather close to each other near $\Omega_{\mathrm{m}} = 0.30$.
 The observations, which are essentially centered on the value by \citet{Farooq13}, extend over a range compatible
 with the two kinds of models. 
 The distinction between them may be possible in the future with accurate data, for now it is still uncertain.
On the whole we conclude that  the observations are in  good agreement with both the  flat $\Lambda$CDM and  scale invariant
models.
  }}

\section{Conclusions}

 The discovery of the accelerated expansion of the Universe
has shown that  the  situation  is like if an interaction of an
unknown nature opposes the gravitation. In this context, the  hypothesis of scale invariance,
which ''naturally'' leads to an acceleration of the expansion,      opens a window on possible interesting 
 cosmological models.  
 The first comparisons of models and observations made so far
{ {\bf{on the distances, the $m-z$ diagram, the expansion rate $H_0$, the dynamical properties of the scale invariant cosmology 
 and the transition from braking to acceleration show  consistent results.  
If true, the hypotheses we made have many other implications in astrophysics and cosmology. 

Thus, for now these cosmological models evidently  need 
 to be further thoroughly  checked with many other possible astrophysical tests in order to confirm or infirm them.
In view of further tests, a point  about methodology needs to be   emphasized:  
to be valid, a test must be internally coherent and not rest on properties or inferences  from the framework of other
 cosmological  models, a point which is not always evident.}}}

\vspace*{3mm}

\noindent
Acknowledgments: I want to express my best thanks to the physicist D. Gachet and Prof. G. Meynet for their 
continuous encouragements.

\allauthors

\listofchanges

\end{document}